\documentclass[11pt,preprint,single]{aastex}
\newcommand{\etal}{{\it et al.} }
\newcommand{\asca}{{\it ASCA} }

\newcommand{\xmm}{{\it XMM-Newton} }
\newcommand{\chandra}{{\it Chandra} }

\newcommand{\rxte}{{\it RXTE} }
\newcommand{\hetg}{{\it HETGS} }
\newcommand{\letg}{{\it LETGS} }
\newcommand{\fekalfa}{{Fe~K$\alpha$} }
\newcommand{\bsax}{{\it BeppoSAX} }

\newcommand{\sila}{Si~{\sc xiv}~Ly$\alpha$ ($\lambda 6.182$) }
\newcommand{\nela}{Ne~{\sc x}~Ly$\alpha$ ($\lambda 12.134$) }
\newcommand{\nila}{N~{\sc vii}~Ly$\alpha$ ($\lambda 24.781$) }
 
\newcommand{\oxla}{O~{\sc viii}~Ly$\alpha$ ($\lambda 18.969$) }
\newcommand{\mgla}{Mg~{\sc xii}~Ly$\alpha$ ($\lambda 8.421$) }

\newcommand{\oxlb}{O~{\sc viii}~Ly$\beta$ ($\lambda 16.006$) }

\newcommand{\oxyseven}{O~{\sc vii} }
\newcommand{\oxyeight}{O~{\sc viii} }
\newcommand{\oxysevenf}{[O~{\sc vii}] }
\newcommand{\oxyseveni}{O~{\sc vii} (i) }
\newcommand{\oxysevenr}{O~{\sc vii} (r) }
\newcommand{\nenine}{Ne~{\sc ix} }
\newcommand{\neninef}{[Ne~{\sc ix}] }
\newcommand{\neniner}{Ne~{\sc ix} (r) }

\newcommand{\feeighteen}{Fe~{\sc xviii} }  
\newcommand{\fenineteen}{Fe~{\sc xix} }
\newcommand{\fetwenty}{Fe~{\sc xx} }

\newcommand{\fetwentytwo}{Fe~{\sc xxii} }
\newcommand{\fetwentythree}{Fe~{\sc xxiii} }
\newcommand{\fetwentyfour}{Fe~{\sc xxiv} }
\newcommand{\fetwentyfive}{Fe~{\sc xxv} }
\newcommand{\mgeleven}{Mg~{\sc xi} }
\newcommand{\mgelevenr}{Mg~{\sc xi} (r) }

\newcommand{\sithirteen}{Si~{\sc xiii} }

\newcommand{\res} {${\it 1s^{2}-1s np}$ }
\newcommand{\resonetwo} {${\it  1s^{2}-1s 2p}$ }
\newcommand{\resonethree} {${\it 1s^{2}-1s 3p}$ }
\newcommand{\resonefour} {${\it 1s^{2}-1s 4p}$ }
\newcommand{\figufspecnhgal}{Fig.~1 }
\newcommand{\figufspecedges}{Fig.~2 }
\newcommand{\figufspeccont}{Fig.~3 }
\newcommand{\figmeglyman}{Fig.~4 }
\newcommand{\figmegbalmer}{Fig.~5 }
\newcommand{\figtriplets}{Fig.~6 }
\newcommand{\figvelprofone}{Fig.~7 }
\newcommand{\figvelproftwo}{Fig.~8 }
\newcommand{\figsed}{Fig.~9 }
\newcommand{\figxstaroverdata}{Fig.~10 }
\newcommand{\figxstarmodel}{Fig.~11 }
\newcommand{\figxstarcontour}{Fig.~12 }
\newcommand{\figcog}{Fig.~13 }

\newcommand{\src}{NGC~4593 }
\newcommand{\ngc}{NGC~4593 }

\newcommand{\mcg}{MCG~$-$6$-$30$-$15 } 

\begin{document}

\title{THE KINEMATICS AND PHYSICAL CONDITIONS OF THE IONIZED GAS IN NGC 4593 FROM CHANDRA HIGH-ENERGY GRATING SPECTROSCOPY}

\author{Barry McKernan\altaffilmark{1},
Tahir Yaqoob\altaffilmark{1,2},
Ian M. George\altaffilmark{2,3},
T. J. Turner\altaffilmark{2,3}}

\slugcomment{Accepted for publication in the Astrophysical Journal} 

\altaffiltext{1}{Department of Physics and Astronomy,
Johns Hopkins University, Baltimore, MD 21218}
\altaffiltext{2}{Laboratory for High Energy Astrophysics,
NASA/Goddard Space Flight Center, Greenbelt, MD 20771}
\altaffiltext{3}{Joint Center for Astrophysics, University of Maryland,
Baltimore County, 1000 Hilltop Circle, Baltimore, MD 21250}


\begin{abstract}
We observed the Seyfert~1 galaxy \ngc with the \chandra high energy 
transmission gratings and present a detailed analysis of the soft X-ray spectrum. We measure strong absorption lines from He-like O, Ne, Mg, Si, H-like 
N, O, Ne, Mg, Si and highly ionized Fe~{\sc xix-xxv}. 
The weighted mean of the offset velocity of the strongest absorption profiles is $-140 \pm 35$ km $\rm{s}^{-1}$. However the individual profiles are consistent with the systemic velocity of \ngc and many profiles hint at the presence of either multiple kinematic 
components or blending. Only the
\nila, \oxla and \mgla lines appear to be marginally resolved. 
We identify a spectral feature at 
$\sim 0.707$ keV with a neutral Fe L edge, which might suggest that there is dust along the line-of-sight to \ngc although, this is not the only interpretation of this feature. A search for neutral O absorption which would reasonably be expected from dust absorption is complicated by contamination of the \chandra ACIS CCDs. Neutral Si absorption, which might also be expected from absorption due to dust is present (though not significantly) in the form of a weak neutral Si edge. The neutral Si column ($<4 \times 10^{17} \ \rm{cm}^{-2}$) corresponding to the Si edge is consistent with the neutral Fe column ($\sim 1.5 \times 10^{17} \ \rm{cm}^{-2}$) from the Fe L edge. We also detect, at marginal signficance, \nila and
\oxysevenr \resonetwo ($\lambda 21.602 $) absorption at $z \sim 0$, 
due to a hot medium in our Local Group. The soft X-ray 
spectrum of \ngc is adequately described by a simple, 
single-zone photoionized absorber
with an equivalent Hydrogen column density of $5.37^{+1.45}_{-0.79} 
\times 10^{21} \ \rm{cm}^{-2}$ and an ionization parameter of 
$\log \xi= 2.52^{+0.06}_{-0.04}$ ergs cm $\rm{s}^{-1}$ although there remain some features which are not identified. Although the photoionized gas
almost certainly is comprised of matter in more than one ionization state 
and may consist of several kinematic components, data with better 
signal-to-noise ratio and better spectral resolution are
required to justify a more complex model. Finally, in emission we detect only 
weak forbidden lines of \neninef and \oxysevenf.

\end{abstract}


\section{Introduction}
\label{sec:intro}
 
The launch of \chandra and \xmm began a new era in the
study of X-ray photoionized circumnuclear gas. High resolution X-ray
spectroscopy with \chandra now allows the gas kinematics to be
studied seriously for the first time, and the detection
of individual absorption and emission lines can now place
very strong constraints on the ionization
structure of the warm, or partially ionized, X-ray absorbing gas in
type~1 active galactic nuclei (AGNs)(e.g. \cite{lee01}; \cite{sako01a}; \cite{kaastra02}; \cite{kaspi02}; \cite{yaqoob02}).

Here we present the results of a 
\chandra High-Energy Transmission Grating (or \hetg --Markert, \etal 1995) 
observation of \ngc, which was simultaneous with an \rxte observation.
\ngc ($z=0.0083$\footnote{Deduced from observations of the 21cm H~{\sc i} line. Uncertainty in this measurement ($\pm6$ km $\rm{s}^{-1}$) is negligible compared to the systematic uncertainty in the \chandra MEG energy scale ($\pm 67$ km $\rm{s}^{-1}$ at 0.5 keV) and also much less than the best MEG spectral resolution ($\sim 280$ km $\rm{s}^{-1}$ at 0.5 keV). Optical observations suggest a range of $z=0.0080-0.0090$, corresponding to a range of $-90$ km $\rm{s}^{-1}$ to $+180$ km $\rm{s}^{-1}$. However, optical observations may be confused by AGN outflow and so are less accurate indicators of source redshift.}, De Vaucouleurs \etal 1991) is a type~I AGN that lies within a barred-spiral host galaxy. \ngc is also relatively luminous ($L_{2-10 \rm \ keV}$ typically $8.6 \times 10^{42} \ \rm ergs \ s^{-1}$, converted from the luminosity given in Reynolds (1997) using $H_{0}=70 \ \rm km \ s^{-1} \ Mpc^{-1}$ and $q_{0}=0$ which we use throughout this paper). In the present paper we focus on the soft X-ray spectroscopy results. 

The paper is organized as follows.
In \S\ref{sec:obs} we present the data and describe the analysis
techniques. In \S\ref{sec:overall} we discuss gross features
of the X-ray spectrum, including the intrinsic continuum form and
interpret the data in the context of historical, lower spectral
resolution CCD data. In
\S\ref{sec:features} we qualitatively
discuss the discrete X-ray spectral features before describing
detailed spectral modeling.
In \S\ref{sec:modelling} we describe in detail the modeling of
the X-ray spectrum using the photoionization code, XSTAR.
In \S\ref{sec:compare} we compare our results with
those from other X-ray observations of \ngc. In \S\ref{sec:dust} we discuss 
the possibility of a dusty warm absorber in \src and we investigate the 
relativistically broadened disk-line scenario (Branduardi-Raymont \etal 2001; Sako \etal 2001b; Mason \etal 2002) as an alternative to the 
dusty warm absorber model. Finally, in \S\ref{sec:results} we summarize our results and in \S\ref{sec:conclusions} briefly state our conclusions. 

\section{Observations and Data}
\label{sec:obs}
 
We observed \ngc
with \chandra (simultaneously with \rxte)
on 2001 June 29
for a duration of $\sim 79$~ks,
beginning at UT 00:54:49.
The \chandra data were reprocessed
using {\tt ciao 2.1.3} and
{\tt CALDB} version 2.7, according to
recipes described in
{\tt ciao 2.1.3} threads\footnote{http://asc.harvard.edu/ciao2.1/documents\_threads.html}.
The \rxte PCA data were reduced using methods described in
Weaver, Krolik, \& Pier (1998), except 
improved background subtraction was utilized
(released 2002, February) and a later version of the spectral response matrix
(v 8.0) was used\footnote{Details of both 
in http://heasarc.gsfc.nasa.gov/docs/xte/xhp\_proc\_analysis.html}.
Only data from layer 1 and from
PCUs 1, 2, 3 and 4 were used (PCU0 has had technical problems in the later
part of the mission).
 
For \chandra
the instrument used in the focal plane of the
High Resolution Mirror Assembly was the \hetg. The \hetg consists of two grating assemblies, a 
High-Energy Grating (HEG) and a Medium-Energy Grating (MEG). Only the summed, negative and positive, first-order \chandra grating spectra were used in our analyis. For further 
discussion of the \hetg and our analysis procedure here, 
see Yaqoob \etal (2003). The mean \chandra total HEG and MEG count rates
were $0.1965 \pm 0.0017$ and $0.4743 \pm 0.0021$ cts/s respectively.
HEG, MEG and PCA 
spectra were extracted over the entire on-time for 
each instrument. This resulted in net exposure times of 78,889 s
for HEG and MEG, and between $\sim 14$~ks and $\sim 70$~ks for
each PCU. The background subtracted spectra from the different PCUs 
were co-added for analysis and the resulting mean count rate was 4.8 ct/s over
the entire PCA band. The excess variances on the hardness ratios 1.0--2.5 keV/(0.6--1.0 keV) and 2.5--7.0 keV/(0.6--1.0 keV) for \chandra data binned at $0.16\AA$ were $0.0011 \pm 0.0012$ and $0.0014 \pm 0.0007$ respectively. Thus, the data are consistent with no spectral variability in soft X-rays during the observation, but there may be some small spectral variation in hard X-rays. The \rxte observation extended
$\sim 20$~ks beyond the \chandra observation but the mean count rate
in that portion of the observation was compatible with the
mean of the strictly overlapping portion. Therefore, for the
sake of improved statistics, the PCA spectrum included
$\sim 10$~ks exposure beyond the \chandra data.
 
For \chandra \hetg we made effective area files (ARF, or {\it ancillary response file}),
photon spectra and counts spectra following the method of 
Yaqoob \etal (2003). Since the MEG
soft X-ray response is much better than the HEG
(whose bandpass only extends down to $\sim 0.8$ keV) we will
use the MEG as the primary instrument but refer to the HEG
for confirmation of features in the overlapping bandpass and
for constraining the continuum.
The MEG spectral
resolution corresponds to FWHM velocities of $\sim 280, 560$, and
3,560~$\rm km \ s^{-1}$ at observed energies of 0.5, 1.0, and 6.4 keV 
respectively.

We did not subtract detector or X-ray background since it is
such a small fraction of the observed counts
($<0.5\%$ for the MEG).
We treated the statistical errors on both the photon and counts
spectra with particular care since the lowest and highest energies
of interest can be in the Poisson regime, with spectral bins often
containing a few, or even zero counts.
We assign statistical upper and lower errors of $1.0 + \sqrt{(N+0.75)}$ and $N(1.0-[1.0-(1/9N)-(1/3\sqrt(N))]^3)$ respectively (from Gehrels, 1986)
on the number of photons, $N$, in a given spectral bin. Statistical
errors were always computed on numbers of photons in final bins,
according to the above prescription for plotting purposes. When fitting the \chandra data, we used the $C$--statistic (Cash, 1976) for finding the best-fitting model
parameters, and quote 90$\%$ confidence, one-parameter statistical errors 
unless otherwise stated. The $C$--statistic minimization algorithm is inherently Poissonian and so makes no use of the errors on the counts in the spectral bins described above. 
  
For the HEG data, the signal-to-noise ratio per $0.02\AA$ bin
is $> 1$ blueward of  $\sim 15\AA$ and 
as high as $\sim 10$ in the $\sim 6-7\AA$ range. 
For the MEG data, in the $\sim 2-25\AA$ band which we examine
here,
the signal-to-noise ratio per $0.02\AA$ bin ranged from $\sim 1$
at $\sim 22.5\AA$ to a maximum of $\sim 14$ at $\sim 7\AA$.
The systematic uncertainty in the energy scale is currently
believed to be $0.0028\AA$ and $0.0055\AA$ for the HEG and
MEG respectively\footnote{http://space.mit.edu/CXC/calib/hetgcal.html}.
For the MEG, at 0.5 keV, 1 keV and 6.4 keV this corresponds
to velocity offsets of $\sim 67$, 133, and 
852~$\rm km \ s^{-1}$ respectively, and about half of these values for the HEG.

In addition, there has been a continuous degradation of the quantum efficiency 
of \chandra ACIS due to molecular contamination. The degradation 
includes an excess absorption due to neutral O \footnote{http://cxc.harvard.edu/cal/Links/Acis/acis/Cal\_prods/qeDeg/index.html}. 
When fitting photoionization models to the data we 
will investigate the effect of this degradation on
model parameters. There is no satisfactory correction available for the gratings but we will estimate worst case effects using a pure ACIS correction, since a grating in front of the CCDs will not make the CCD degradation worse. Indeed a grating may improve the situation by adding extra energy information. We will show in \S\ref{sec:modelfits} that omitting the correction only affects the inferred intrinsic continuum and does not affect the important physical parameters of absorption models.

\section{Overall Spectrum and Preliminary Spectral Fitting}
\label{sec:overall}

We used XSPEC v11.2.0 for spectral fitting to the HEG and MEG spectra
in the 0.8--5 keV and 0.5--5 keV bands respectively.
These energy bands will be used in all the spectral fitting in the
present paper, in which we concentrate on 
features in the soft X-ray spectrum (less than $\sim 2$ keV). The harder spectrum out to $\sim$ 15 keV, including the
Fe-K emission line and Compton-reflection 
continuum (using simultaneous
\rxte data in addition to the \chandra data) will be discussed elsewhere. 
However, we will utilize a portion of the \rxte PCA spectrum to constrain the
continuum for fitting the soft X-ray \chandra data and use the $\chi^{2}$ 
statistic for minimization. 

Since we convolve all
models through the instrument response before comparing with the
data, all our model line-widths are intrinsic values and do
not need to be corrected for the instrument broadening. This
contrasts with the analysis method in some of the literature
on results from grating data in which models are fitted directly to the
data and {\it then} corrected for instrument broadening 
(e.g. Kaspi \etal 2002). First, we show how the 0.5--5 keV 
MEG data compare to a
simple model consisting only of a 
single power law
and Galactic absorption. For the latter, we use a value of
$1.97 \times 10^{20} {\rm cm^{-2}}$ (Elvis \etal 1989) throughout this
work. Later in this paper we show that the ACIS degradation discussed above does not effect our results. Therefore we do not let the Galactic column vary in spectral fitting since its systematic error is much less than the effects of ACIS degradation. \figufspecnhgal (a) shows the MEG photon spectrum of \ngc 
binned at $0.32\AA$ (which is comparable with \asca resolution) 
with the model overlaid. The power-law index was allowed to vary 
in this fit. \figufspecnhgal (b) shows the ratio of the data to this model. 
This simple power law is clearly a poor fit to the data, with considerable
complexity apparent, both in the continuum, and in terms of
discrete absorption features.
\figufspecnhgal (b) clearly shows a soft
X-ray excess, rising up from the hard power law
(modified by Galactic absorption), below $\sim 0.7 \ {\rm keV}$.
We caution that there may still be broad residual uncertainties of
$\sim 30\%$ or more in the calibration of the MEG effective area
at the lowest energies
\footnote{http://space.mit.edu/CXC/calib/hetgcal.html}. The instrumental effective area is very small at these energies, although residuals at $<0.6$ keV probably coincide with the complex instrumental O~{\sc{i}} edge which is spread over $\sim$ 10 eV around $\sim 0.53$ keV.

Observations of \ngc in 1994 with \asca, as summarized in Reynolds (1997) 
and George \etal (1998) also indicated a soft excess present below 
$\sim 0.7$ keV, relative to an extrapolation of the best-fitting 
hard X-ray power-law. 
A soft excess has  also been observed
by \xmm and \chandra \letg (Kaastra  \etal 2001), as well as  by
\bsax (Guainazzi \etal 1999). This would suggest that at
least part of the soft excess we observed with \chandra \hetg
is real. The low-energy degradation of the CCDs mentioned in \S\ref{sec:obs} is in
fact in the sense that any soft excess will actually be
larger when the correction is taken into account. 

The origin of the soft excess is unclear. It may be the tail end of 
some kind of soft thermal emission (or Comptonized soft
thermal emission), possibly from an accretion disk 
(e.g. see Piro, Matt, \& Ricci (1997), 
and references therein). Any relativistically-broadened
\oxla emission that is present could also contribute to the
soft excess (e.g. Branduardi-Raymont \etal 2001; Turner \etal 2001).
Since the soft excess appears only in the 0.5--1 keV band
of our data, we do not have enough information to constrain
its origin, and sophisticated modeling of it is not warranted.
Therefore, in the remainder of this paper we model the
0.5--5 keV intrinsic continuum with a broken power law.
When the data are modeled with this continuum, there is
still considerable structure in the soft X-ray spectrum,
characteristic of the presence of a photoionized, or `warm' absorber.

Now, the warm absorber in \ngc and 
other AGN when observed with CCDs (such as those on \asca)
has often been modeled simply with absorption edges
due to \oxyseven and \oxyeight since the lower CCD spectral
resolution did not usually warrant more sophisticated models.
In order to directly compare the new \chandra data with
previous observations we modeled the MEG data for \ngc with
two absorption edges, Galactic absorption,
and the intrinsic continuum described 
above. 
Here we used the \rxte PCA data to help constrain the continuum.
First, we modeled the HEG, MEG, and PCA data simultaneously,
the latter data only covering the 3--8 keV band in order to
avoid the Compton-reflection continuum. We had to include
an additional Gaussian model component (with center energy,
width and intensity allowed to vary) to model the Fe-K line.
Since we are not concerned with the Fe-K line in this paper
we then take the hard power-law index 
($\Gamma =1.794^{+0.013}_{-0.019}$) from this fit and {\it use $\Gamma = 1.794$ in all subsequent model-fitting in this
paper}. If we allow $\Gamma$ to vary in our best model (two absorption edges, a warm absorber and a broken power-law , see \S\ref{sec:results} below) fit to the \chandra MEG data binned at 0.16$\AA$, we find a best-fitting value of $\Gamma=1.803 \pm 0.017$ which is consistent with the fixed value above, so fixing $\Gamma$ does not affect our results or conclusions. Since we will only be using \chandra data below 5 keV from 
this point on, we do not have to worry about the spectral 
fits becoming unstable due to too much interplay between the
two parts of the broken power-law index and the Fe-K line model.
This will be dealt with in future work.

In order to measure the absorption-edge
parameters, we then modeled the 0.5--5 keV
MEG data only, fixing the
hard photon index at 1.794,
and allowed the energies and optical depths (at threshold) of
the two absorption edges to vary.
The best-fitting model overlaid on the MEG photon spectrum
is shown in \figufspecedges. Although a spectrum with
bin size $0.04\AA$ was used for the spectral fitting,
the plot shows the data binned at $0.32\AA$ since this
is closer to the resolution of CCD data.
The best-fitting parameters obtained from this
model were $0.708 \pm 0.003$ keV and $0.868^{+0.007}_{-0.010}$ keV
for the edge threshold energies. 
For the the optical depths at threshold
we obtained $\tau = 0.18\pm 0.04$ and 
$0.14^{+0.03}_{-0.04}$ respectively.
All quantities refer to the source rest frame.
The best-fitting soft X-ray photon index was $2.26\pm 0.05$ and the
break energy was $1.03^{+0.04}_{-0.03}$ keV. 

The edge at 0.868 keV 
is in good agreement with the expected energy of 
the \oxyeight edge at 0.871 keV. However, the edge at 0.708 keV is redshifted 
by $\sim 12,600$ km $\rm{s}^{-1}$ from the expected energy of the \oxyseven 
edge (0.739 keV). Rather, this edge energy agrees extremely well with the 
expected energy of a neutral Fe L3-edge at 0.707 keV (as observed, for 
example, by Lee \etal (2001) in the warm absorber of \mcg). 
Furthermore, adding a \emph{third} edge to the continuum model, with a 
threshold energy fixed at that expected for \oxyseven, does not
improve the fit statistic significantly. The threshold optical depth of the 
\oxyseven edge in this case is $<0.03$ at the 
one-parameter, $90\%$ confidence level.
We note that since the spectrum is so complex around
the regions of the \oxyseven and \oxyeight edges,
optical depths derived from simple models should
be interpreted with caution. 
We can compare the threshold optical depths with those measured from the 1994 
January 4 \asca observation of \ngc (Reynolds 1997): $0.26\pm 0.04$ 
(ascribed to \oxyseven) and $0.09^{+0.04}_{-0.03}$ (\oxyeight). 
The \chandra and \asca \oxyeight edge depths are in marginal agreement
but the \oxyseven depths are completely incompatible. It seems that, due to 
the limited energy resolution of CCD data, the Fe-L3 edge was misidentified
with an \oxyseven edge, not only in \ngc but possibly also for other AGN. The high resolution of \chandra \hetg indicates that this edge \emph{cannot} be due to \oxyseven, and previous \asca data from AGN needs to be re-examined in this context. The optical depth that we measure for the Fe-L3 edge is in marginal
agreement with 
that ascribed to the \oxyseven edge by Reynolds (1997). 
The \asca SIS (4-CCD mode) observation of \ngc had an energy resolution (FWHM) 
of $\sim 140$ eV at $\sim 1.0$ keV at the time of the observation, 
and therefore could not distinguish between an \oxyseven edge 
and a neutral Fe edge. 
However, an Fe-L edge for the feature at 0.7 keV is not the only 
interpretation (see \S{\ref{sec:dust}}). High resolution observations with \chandra are also hinting at X-ray absorption fine structure (XAF) around absorption edges (Lee \etal 2002). The apparent Fe-L edge in our data extends over $\sim 5$ eV, which is $\sim 3$ times the FWHM resolution of \chandra at this energy. The S/N of our data and the FWHM resolution of \chandra are not adequate to measure the XAF structure around the Fe-L edge, which in any case, will be complicated by \oxyseven \res ($n>4$) absorption features.

\section{Spectral Features and Kinematics}
\label{sec:features}

\subsection{Absorption and Emission Lines}
\label{sec:abslines}
In order to best illustrate different characteristics of
the spectrum we display the spectral data in several
different ways. In \figufspeccont we show the MEG photon
spectrum between 0.47--2.1 keV, as a function of observed energy. 
The data and the models in \figufspeccont are all absorbed
by the Galactic column density.
An important point to realize from \figufspeccont is that
in some regions the spectrum is so complex that it is
difficult to distinguish absorption from emission. For
example, two emission features may be present between 1.0--1.5 keV
according to the data/model ratio plot of \figufspecedges. However, from \figufspeccont
it is clear that the peak flux of both features
coincides with the inferred intrinsic continuum, implying that the apparent
feature is simply due to the presence of broad absorption
complexes on either side of it. In addition there is a sharp variation 
in the MEG effective area (at $\sim 0.85$ keV) and any uncertainties in the 
response could be manifested as an apparent spectral feature 
(see \S\ref{sec:modelling} for further discussion).

In \figmeglyman and \figmegbalmer we show the MEG spectrum, 
this time as a function of wavelength in the source rest-frame, 
redward of $2\AA$. These spectra have bin sizes of $0.02\AA$, approximately
the FWHM MEG spectral resolution ($0.023\AA$).
In \figmeglyman we overlay the Lyman series (blue)
for H-like N, O, Ne, Mg, Si, S, Ar and the corresponding He-like 
triplets (red) that lie in the 2-25$\AA$ range.
In \figmegbalmer we overlay the He-like
resonance series lines (blue) to the $n=1$ level for the same elements. In addition, in \figmegbalmer we indicate L-shell transitions of \fenineteen (red) in the $2-25\AA$ range\footnote{http://www.uky.edu/~peter/atomic/} to illustrate just some of the blending of spectral features that can arise due to absorption by highly ionized Fe.

Referring to Figs. 3 to 5, we 
see that the strongest, clear discrete spectral features detected are
absorption lines due to \oxla, \oxlb, \neniner \resonetwo 
($\lambda 13.447$), \neniner \resonethree ($\lambda 11.549$),
\nela, \mgla, \sithirteen (r) \resonetwo ($\lambda 6.648$). Also included amongst 
the strongest discrete features are transitions (and blends of transitions) which 
we identify with highly ionized Fe, typically L-shell transitions in \fetwenty--\fetwentyfive. Weaker absorption lines are also apparent from the spectra
 such as \nila, \oxyseven (r) \resonetwo ($\lambda 21.602$) and higher-order 
Lyman and (He-like) resonance absorption lines of O and Ne (there is evidence 
for transitions up to $n=4$ in each case; see Figs. 4 and 5). 
Some of these weaker features may be blended with \fetwentytwo - 
\fetwentythree transitions. Higher signal-to-noise X-ray spectroscopy (e.g. of NGC~3783, Kaspi
\etal 2002) has shown that blending of absorption features
considerably complicates the spectrum. In particular, transitions of \fetwentythree could be blended with the absorption 
troughs at $\sim 10.65$ and $11.0\AA$. Similarly, transitions of 
\fetwenty and \fetwentytwo could be blended with troughs at $9.15\AA,12.8\AA$, 
and $11.75 \AA$ respectively. We note that such a wide range of observed 
ionization states (from N~{\sc vii} to \fetwentyfive) 
suggests that there is more than one ionization component of the warm 
absorber. We shall return to this point in the discussion of photoionization 
modeling in \S\ref{sec:modelling}.

We also identify absorption features due to 
\oxysevenr \resonetwo ($\lambda 21.602 $) and \nila in the observed frame, 
at $z=0$, 
indicating absorption along the line of sight due to a hot medium in our Galaxy 
or local group (see also Figs. 3, 4 \& 5). The separations of the lines at $z=0$ 
and $z=0.0083$ is 2490 km $\rm{s}^{-1}$, which is easily resolved by \chandra.
Strong local absorption by \oxysevenr \resonetwo ($\lambda 21.602 $) and 
\neniner \resonetwo ($\lambda 13.447$) and weaker local absorption by \oxla 
and \oxysevenr \resonethree ($\lambda 18.628 $) has also been observed by Nicastro 
\etal (2002) in the direction of the blazar PKS 2155-304. Nicastro \etal (2002)
suggest that the local absorption is produced in a low density intergalactic 
plasma, collapsing towards our Galaxy. It is likely that the $z=0$ absorption
features \oxysevenr \resonetwo ($\lambda 21.602 $) and \nila in \ngc are due to the same gas. The strengths of the lines, although only marginally significant, are comparable with the EW$\sim 0.2-0.3$ eV detected by Nicastro \etal (2002).

Apart from \fekalfa (which will be discussed elsewhere), we also 
identify weak \oxysevenf and \neninef emission features. 
Table~1 shows that \oxysevenf and \neninef are 
redshifted but the errors are consistent with systemic velocity. 
The corresponding intercombination and resonance lines of the \oxyseven 
and \nenine He-like triplets are not 
detected. The relative prominence of forbidden lines compared to the 
intercombination and resonance lines indicates an 
origin in a photoionized plasma (e.g. Porquet \& Dubau 2000). 
\figtriplets shows close-ups of the wavelength 
regions spanning 
the \oxyseven and \nenine He-like triplets (with the positions of the 
forbidden (f), intercombination (i) and resonance (r) lines marked in red). 
The ratio of the forbidden and intercombination line fluxes (R=f/i) constrains
the electron density (e.g. see Porquet \& Dubau 2000). An estimate of R 
from the flux at the i,f wavelengths in \figtriplets yields R$<$3 for \nenine 
and R$<$10 for \oxyseven. 
These ratios imply $n_{e}< 2 \times 10^{12} \ {\rm cm^{-3}}$ 
and $n_{e}<8\times 10^{10} \ {\rm cm^{-3}}$ respectively, for a relative 
ionic abundance of the H-like and He-like ions of $\leq 10^{3}$ (see Fig.9 of 
Porquet \& Dubau 2000). If blending is significant in either of these triplets, then the measured line intensities would be upper limits only, thereby yielding a potentially significant systematic error to our estimates of R and therefore $n_{e}$.

The presence of broad, bound-free absorption features
and many narrow features which may be blended,
makes it impossible to define an observed continuum
for much of the spectrum,
and therefore it is 
difficult to measure meaningful equivalent widths of
even the strongest narrow absorption features.
Rather than try to measure equivalent widths 
of {\it all} candidate features, and compare
with model predictions, we will take the physically more
direct approach of fitting photoionization models to the data
(\S\ref{sec:modelling}). First we give the measured energies and equivalent widths of the strongest absorption lines. The measurements were made using the
broken power law plus two-edge model as the baseline
spectral model (see \S\ref{sec:overall}) and Gaussians
to model the absorption lines. The continuum and edge parameters
were frozen at the best-fitting values given in
\S\ref{sec:overall}, and the data were fitted over
a few hundred eV centered on
the feature of interest. Allowing the edge parameters to vary did not change our results. The fitted range of the data was
extended down to 0.47 keV in order to measure the
parameters of \nila. 
Initially the {\emph{intrinsic}} width of each Gaussian
was frozen at less than the instrumental resolution but then
allowed to vary in order to investigate whether the lines
were resolved. The results for the
measured energies, apparent velocity shifts relative
to systemic, equivalent widths, intrinsic widths and the improvement in the fit-statistic, of these strongest absorption features 
are given in Table~1. Note that the quoted equivalent
widths are relative to the {\it intrinsic} continuum 
(corrected for Galactic absorption). 
Again, we caution that the measured equivalents widths are subject
to uncertainties in the continuum, line-blending, and any complexity
in the line profile. It is instructive to compare
line strengths with a physical model, so also in
Table~1 are the predicted equivalent widths from the
best-fitting photoionization model discussed in \S\ref{sec:modelling}. 

\subsection{Absorption-Line Velocity Profiles}
\label{sec:profiles}

Table~1 shows that most of the measured absorption lines
are blueshifted, with Gaussian-fitted centroids ranging
from near systemic velocity to $\sim -$310 km $\rm{s}^{-1}$. The 
exceptions are \neniner \resonetwo ($\lambda 13.447$) and \fetwenty ($\lambda 12.817 $).
The former is redshifted, but the measurement is consistent with 
systemic velocity and
the latter is likely blended with nearby \fetwenty transitions (see below).
Some of the velocity profiles of the absorption features in Table~1 appear 
to be more complex than Gaussian, as can be seen from an inspection of 
the profiles in Fig.~7 and Fig.~8. These velocity profiles
were obtained from combined HEG and MEG data except for those transitions at 
$>17\AA$, which were determined using MEG data only, 
since they are out of the HEG bandpass.
The profiles are centered with zero at 
the systemic velocity of \ngc, with positive and negative velocities
corresponding to redshifts and blueshifts respectively. 
Superimposed on
the data are velocity spectra from the best-fitting photoionization model 
which will be discussed in detail in \S\ref{sec:modelling}. 
 
\figvelprofone~(a) shows the velocity profile
centered on the \oxla absorption transition in the \ngc
rest-frame. A broad absorption trough can be seen between
$\sim -$600 km $\rm{s}^{-1}$ and systemic, centered
around $\sim -$300 km $\rm{s}^{-1}$. 
\figvelprofone~(b) is centered around the \nela transition and also exhibits a 
broad 
absorption profile from $\sim -$300 km $\rm{s}^{-1}$ to $\sim$ +100 km 
$\rm{s}^{-1}$.
\figvelprofone~(c) shows the velocity profile for \mgla, which exhibits a 
broad absorption profile from $\sim -$500 km $\rm{s}^{-1}$ to $\sim$ +300 
km $\rm{s}^{-1}$. \figvelprofone~(d) shows the velocity profile for \sila. 
The \sila profile appears to be broader than the others, with a flat base between 
$\pm$ 600 km ${\rm s^{-1}}$. 
There is also apparent emission at +900 km $\rm{s}^{-1}$ relative to systemic for 
\sila. This is problematic since \sila 
emission at +900 km $\rm{s}^{-1}$
would be difficult to reconcile with the small redshifts (consistent with 
systemic) of the \oxysevenf and \neninef emission features. All four 
profiles in \figvelprofone have widths that agree with the formal 
upperlimits in Table~1 ($\sim$ few hundred to $\sim$ 1000 km ${\rm s^{-1}}$ FWHM for all profiles, except the more complex and broader \sila profile which is $\sim 2000$ km ${\rm s^{-1}}$ FWHM ). Formal results from Gaussian fits indicate that both \oxla and \mgla are marginally 
resolved (see Table~1).

\figvelproftwo shows the velocity profiles of some of the strongest resonance 
absorption features and the profile of a transition in highly ionized Fe. 
Superimposed are velocity spectra from the best-fitting photoionization model 
(solid lines) and from the 99$\%$ upper limits on this model (dashed lines, superimposed in \figvelproftwo~(b,c) only) (see 
\S\ref{sec:modelling} for further details).
\figvelproftwo~(a) is centered on the \fetwenty ($\lambda 12.817$) transition. 
Absorption is strong and spans $\sim -$300 km $\rm{s}^{-1}$ to $\sim$ +500 km 
$\rm{s}^{-1}$. The profile appears deeper on the redward side of systemic. This may
 be due to a second absorber kinematic component, or due to blending with a 
one or more Fe L-shell transitions. Formally, from Gaussian fits, this Fe blend is marginally resolved (see Table~1). \figvelproftwo~(b) is centered on the \neniner 
\resonetwo ($\lambda 13.447$) transition. The absorption trough spans 
$\sim -$100 km ${\rm s^{-1}}$ to $\sim$ +300 km ${\rm s^{-1}}$. 
Also noticeable in \figvelproftwo~(b) is 
the sharp absorption feature at $\sim$ +1500 km ${\rm s^{-1}}$. This probably 
corresponds to a blend of \feeighteen ($\lambda 13.503$) and
\fenineteen ($\lambda 13.504$) L-shell transitions, 
which are at $\sim$ +1300 km ${\rm s^{-1}}$ offset in velocity space from 
\neniner \resonetwo ($\lambda 13.4473$). 
\figvelproftwo~(c) shows the absorption velocity profile of \mgelevenr 
\resonetwo ($\lambda 8.421$). 
The profile minimum lies around $\sim -$230 km ${\rm s^{-1}}$, and the profile 
which spans $\sim -$600 km ${\rm s^{-1}}$ to systemic velocity, 
appears slightly broader on the blueward side, suggesting an unresolved 
kinematic 
component or an unresolved blend with (most likely) \fetwenty L-shell 
($\lambda 9.166-9.167$) transitions. 
Finally, \figvelproftwo~(d) is centered on the \sithirteen (r) \resonetwo ($\lambda 6.648$) transition which spans $\sim -$300 
km ${\rm s^{-1}}$ to systemic velocity. 

Just as for the equivalent
widths, one should not interpret apparent velocity
shifts too literally because,
in addition to uncertainty in the continuum, 
the velocity profiles are complex and in fact can be different
for different ions. The complexity may either be
intrinsic or due to contamination from other
absorption and/or emission features. Emission filling in the red side of absorption features can lead to saturated features appearing non-saturated (Kaspi \etal 2002). If all the lines listed in Table~1 were in fact P Cygni, the estimated outflow velocities in Table~1 would gain a systematic error of about half of the listed outflow velocities. Intrinsic complexity could also be due, for example,
to the presence of several kinematic
components, making different contributions to the overall
profile. However, the fact that most features are consistent with systemic 
(see Table~1) is significant and those features that deviate strongly from 
systemic are likely affected by spectral complexity. 

\section{Photoionization Modeling}
\label{sec:modelling}

We used the photoionization code XSTAR 2.1.h to generate several grids of
models of emission and absorption from photoionized gas in
order to directly compare with the data. We used the default
solar abundances in XSTAR (e.g. see Table~2 in Yaqoob \etal 2003).

\subsection{The Spectral Energy Distribution}
\label{sec:sed}

First we constructed a spectral energy distribution (SED) for \ngc, as 
follows. Average radio, infrared (IR), optical and ultraviolet (UV) fluxes 
were obtained from Ward \etal (1987) and NED\footnote{NASA Extragalactic Database, http://nedwww.ipac.caltech.edu/}. 
The X-ray region of the SED is constructed from the 
intrinsic continuum derived from our \chandra \hetg data (described below). 
Flux measurements obtained from historical data 
(Ward \etal 1987, NED) are not in general contemporaneous. 
Nevertheless, the \ngc SED appears to
exhibit a peak in the IR-UV spectral region.
Due to the large spectral variability however, the flux levels in this spectral
region can span up to an order of magnitude at a given frequency.
We attempted to crudely account for variability in 
this region of the \ngc SED by generating two different SEDs. For the 
first SED 
(SEDa), we chose the highest envelope of the measurements in the IR-UV region
 from Ward \etal (1987). SEDa exhibits a prominent IR-UV bump as is evident in 
\figsed (SEDa is the solid line). We 
then simply connected the optical/UV endpoint of the IR-UV bump at 
$\sim 10^{15}$ Hz to the X-ray portion of the SED with a straight line in 
log-log space. The second SED (SEDb) was generated simply by 
connecting the highest energy IR data point at $\sim 10^{13.9}$ Hz 
from Ward \etal (1987) to the X-ray portion of the SED, such that SEDb
 does not have a prominent bump in the IR-UV region. SEDb is 
represented in \figsed by filled circles connected by a dashed curve. By 
constructing these two SEDs, we could investigate the effect of 
reasonable uncertainties in the IR-UV region on the X-ray photoionization 
models.
Also shown for comparison in \figsed is the lower envelope of the 
measurements in the IR-UV region from Ward \etal (1987) (dash-dot line).

To derive the X-ray portion of the SED, we fitted the data with a broken power 
law continuum and three absorption edges (as discussed in 
\S\ref{sec:overall} above). The 0.5 keV intrinsic model flux was then simply joined onto the last 
UV (SEDa) and IR (SEDb) points of the lower-energy part of the SED by a straight line in 
log-log space. The hard X-ray power law was extended out to 500 keV. Note that we did not include any correction for the ACIS low energy CCD degradation as there is no satisfactory correction for grating data. Grids of
photoionization models (see \S\ref{sec:model} below) were then generated 
using this SED. Also shown in \figsed is the `mean AGN' SED derived
by Matthews \& Ferland (1987) normalized to the ionizing luminosity of the 
\ngc SEDa.

\subsection{Model Grids}
\label{sec:model}

The photoionization model grids used here are two-dimensional,
corresponding to a range in values of
total Hydrogen column density, $N_{\rm H}$,
and the ionization
parameter, $\xi = L_{\rm ion}/(n_{e} R^{2})$.
Here  $L_{\rm ion}$ is the ionizing luminosity in the range 1 to 1000
Rydbergs,
$n_{e}$ is the electron density and $R$ is the distance of the
illuminated gas from the ionizing source.
The grids were computed for equi-spaced intervals
in the logarithms of $N_{\rm H}$ and  $\xi$, in the ranges $10^{20-22} \ {\rm cm^{-2}}$
and $10^{1.5-4.0}$ erg  cm ${\rm s^{-1}}$ respectively. From SEDa, we deduce that $L_{\rm ion} \sim 7\times 10^{44}$ ergs $\rm{s}^{-1}$ and therefore the lowest values of $n_{e},\log \xi$ in the grid yield $R \sim 
150$ pc and the highest values of $n_{e},\log \xi$ in the grid yield $R \sim 
7$ light-hours. This range in $R$ in the grid spans the expected location of the warm absorber. We computed grids with $n_{e}$ in the range
$10^{2-11} \ {\rm cm^{-3}}$. For XSTAR models of the 
\emph{absorber}, we confirmed that 
results from fitting the photoionization models to the X-ray
data were indistinguishable for densities in the
range $n_{e}=10^{2} \ {\rm cm^{-3}}$ to $10^{11} \ {\rm cm^{-3}}$. 
Hereafter we will use $n_{e}=10^{8} \ {\rm cm^{-3}}$ 
unless otherwise stated. Different density models will be important when 
we come to model the emission lines. 
 
All models
assumed a velocity turbulence ($b$-value) of $1$~$\rm km \ s^{-1}$, and 
since XSTAR broadens lines by the greater of the
turbulent velocity and the thermal velocity, it is
the latter which is relevant for these models.
As the turbulent velocity increases, the equivalent
width of an absorption line increases for a given column
density of an ion. 
Therefore, to calculate model equivalent widths
correctly one must know what the velocity width is
(note that the $b$-value could represent all sources of
line-broadening). We cannot simply extend the XSTAR model
grids into another dimension (velocity width), and deduce a
$b$-value directly from model-fitting, because we are limited by the
finite internal energy resolution of XSTAR.
The widths of the energy bins
of the XSTAR model grids are in the range
$\sim$ 200--600 km ${\rm s^{-1}}$.
Our choice of velocity width ($b$-value corresponding
to the thermal width) ensures that the
energy resolution of XSTAR is the limiting factor and not
the $b$-value itself. We will then need to take a rather
more complicated approach to model-fitting (described in 
\S\ref{sec:modelfits} below) and deduce a b-value appropriate for the data. 

\subsection{Model-Fitting}
\label{sec:modelfits}

We proceeded to fit the MEG and HEG spectra simultaneously (binned at
$0.04\AA$, which is approximately the upper limit on the FWHM of most of the strong absorption features in Table~1) , with all
corresponding model parameters for the two instruments 
tied together, except for
the relative normalizations. The intrinsic continuum was
a broken power law (the hard photon index again fixed at 1.794)
modified by absorption from photoionized gas (derived from SEDa, which 
has a prominent bump in the IR-UV region, see 
\S\ref{sec:sed}) 
and the Galaxy, as well as an 
absorption edge due to neutral Fe in the rest-frame of \ngc.
Excluding overall normalizations, there were a total
of seven parameters, namely the column density of the
warm absorber, $N_{H}$, the ionization parameter, $\xi$ and the redshift $z$ 
of the warm absorber, the energy E and 
maximum optical depth $\tau$ of the neutral Fe absorption edge, 
the intrinsic soft photon index, $\Gamma_{2}$ and 
the break energy of the broken power-law continuum, $E_{\rm B}$.

Our model-fitting strategy is more complicated than 
simply comparing the XSTAR model spectra with the data since
the XSTAR spectra assume a certain line width ($b$-value), which
itself is limited by
the finite internal energy resolution of XSTAR.
We chose $b= 1 \ \rm  km \ s^{-1}$ in order that the energy resolution of XSTAR
is the limiting factor. The result is that the equivalent
widths of lines in the XSTAR spectra need to be 
calculated more rigorously
using a line width ($b$-value) appropriate for the data.
On the other hand, the ionic columns in the XSTAR models are robust from
this point-of-view and do not depend on the velocity width.
Therefore, after finding the best-fitting ionization parameter
and warm-aborber column density
from a global fit using the model described above, in \S~\ref{sec:comparison} 
we will deduce a $b$-value from a curve-of-growth analysis, using
ionic column densities from the best-fitting global model. We will
then  
construct detailed model profiles for each of the detected
absorption lines.

The best-fitting parameters derived from the XSTAR global model
fits were
$N_{H} = 5.37^{+1.45}_{-0.79}\times 10^{21} \ {\rm cm^{-2}}$ and
$\log{\xi} = 2.52^{+0.06}_{-0.04}$ ergs cm ${\rm s^{-1}}$. The broken
power-law model corresponding to this best fit had a break energy of
$E_{\rm B}=1.07^{+0.08}_{-0.07} \ {\rm keV}$ and a soft power-law 
photon index of
$\Gamma_{1} = 2.27^{+0.08}_{-0.07}$, values which are consistent with the
SED that was used as input to the models. The absorption edge had an energy of
$E=0.708\pm 0.002$ and a optical depth of $\tau=0.19 \pm 0.04$. The photoionized absorber was offset by $-$140 $\pm$ 35 km $\rm{s}^{-1}$ from 
the systemic velocity of \ngc, the weighted mean 
of the values for velocity offsets of absorption features in Table~1, excluding the \fetwenty ($\lambda 12.817$) transition, which is skewed, likely due to blending. We found that a photoionization model based on
SEDb (the SED without a prominent bump in \figsed) yielded best-fit parameters within the $90\%$ errors given
 above ($\Delta C \sim 4$ for no additional parameters). Only a small difference in results between SEDa and SEDb should be expected since the relevant ionizing continuum is mainly in the X-ray band. Since there are no 
significant differences between the photoionization model fits corresponding 
to each SED, hereafter all photoionization model fits shall refer to SEDa (which has a prominent bump in the optical/UV frequency range). Table~2 shows the ionic column densities predicted by our best-fitting model for a range of ions.

\figxstaroverdata shows the best-fitting model
folded through the MEG response and overlaid
onto the MEG counts spectrum.
The lower panel of \figxstaroverdata
shows the ratio of the
data to the best-fitting model. \figxstarmodel shows the actual
best-fitting XSTAR model before folding through the 
instrument response.
\figxstarcontour shows the 68$\%$, 90$\%$, and 99$\%$ joint
confidence contours of $\log{\xi}$ versus $N_{H}$.
Overall, the fit
is good and it is already apparent that
the best-fitting parameters which describe the
overall spectrum also give good fits to the
most prominent absorption lines and edges.
This can also be seen in \figufspeccont which shows
the best-fitting model overlaid on the photon spectrum,
and in \figvelprofone and \figvelproftwo which show close-ups of the data
and model (in velocity space),
centered on particular atomic transitions. In \figufspeccont, some absorption features are discrepant with the model predictions, most notably \sithirteen (r) \resonetwo ($\lambda 6.648$). The data and model disagree here for two reasons; firstly, because the $b$-value used by XSTAR is not appropriate for the individual absorption features, as discussed above. Secondly, the data in \figufspeccont are binned at $0.08\AA$ which has combined the two minima in \figvelprofone~(d) at $\sim -200$ km $\rm{s}^{-1}$ (which we identify with \sithirteen (r) \resonetwo ($\lambda 6.648$) and $\sim -1000$ km $\rm{s}^{-1}$ (unidentified) into a single deep feature. 

The wavelength accuracy of the
XSTAR atomic database is 5 m$\AA$, which corresponds to a velocity 
offset of 120 km $\rm{s}^{-1}$ at 1 keV. This offset is marginally less than 
the offset due to uncertainty in the MEG energy scale (5.5m$\AA$). 
We mentioned in \S\ref{sec:overall} that the 2.0--2.5 keV region
in the \chandra spectra suffers from systematics as large as
$\sim 20\%$ in the
effective area due to limitations in the calibration of
the X-ray telescope absorption edges in that region
\footnote{http://asc.harvard.edu/udocs/docs/POG/MPOG/node13.html}.
We therefore omitted the  2.0--2.5 keV regions of the \chandra spectra
during the spectral fitting so that the 
fits would not be unduly biased.
These and other instrumental features can be seen
in the upper panel of \figxstaroverdata.
All the MEG data in the
0.5--5 keV band are shown here, although the 2.0--2.5 keV
region was omitted during spectral fitting. Other regions
which have sharp changes in the effective area also
show notable mismatches between data and model, but 
these are not as bad as in the 2.0--2.5 keV region. 

The ACIS CCDs have been undergoing a continuous low-energy
QE degradation, probably due to absorption by a
cumulative build-up of
contaminants. Unfortunately, at the time of writing, there is no
correction available for grating data. There is however,
a model of the contamination pertinent for pure ACIS data
(without the gratings). We can use this model as a `worst case'
to estimate the effects on our photoionization modeling and
conclusions. Thus, we repeated the spectral fitting described
above with the addition of the ACIS QE degradation model
in XSPEC v11.2 ({\tt acisabs}), using the defaults
for absorption by the contaminants C, H, O, and N in the
ratio 10:20:2:1 by number of atoms.
Remarkably, we found that the best-fitting $N_{H}$, $\log{\xi}$
and their associated statistical errors, were
essentially indistinguishable from those obtained
without the contamination model. What changed was
the soft photon index, which increased to $3.04^{+0.04}_{-0.05}$.
Even the break energy remained essentially the same.
Thus, our photoionization modeling is driven by the
discrete, narrow absorption features in NGC~4593 and is
therefore robust to broadband calibration uncertainties. To determine why the degradation model did not change $N_{H}$ or $\xi$, we simulated data consisting of an absorbed power--law continuum with and without degradation respectively. We then fitted the data with and without the {\tt acisabs} model. We found that when the simulated spectrum with no degradation was fitted with {\tt acisabs}, $N_{H}$ did not increase, but the continuum steepened. This is because the degradation yields deep absorption edges which are not in the simulated data. We conclude that the real data have much less degradation than is assumed by the {\tt acisabs} degradation model.

\subsection{Curve-of-Growth Analysis}
\label{sec:cog}
In order to determine the correct velocity width ($b$-value) of absorption 
lines for our models, we calculated curves-of-growth for different values of $b$ 
(see \figcog). The theoretical curves-of-growth in \figcog indicate that the measured equivalent widths of 
$\sim 10^{-2.8} \lambda$ are most consistent with $b \sim 100-200$ km $\rm{s}^{-1}$. We also plotted the measured equivalent widths from Table~1 (converted to 
Angstroms) as a function of ionic column density (from the best-fitting XSTAR 
model described above) and overlaid these on the curves-of-growth (\figcog).
For solar abundances, most of the measurements 
are consistent with $b \sim 200$~$\rm km \ s^{-1}$ within errors. Note that the predicted equivalent widths
for $b=200$~$\rm km \ s^{-1}$ are shown in Table~1
so that they can be compared directly with the observed 
equivalent widths. 

All of the plotted features in \figcog are consistent with the curves of
growth except for \oxyseven (r) \resonetwo ($\lambda 21.602$). The
disagreement with the curves of growth may be indicating a second
component for the warm absorber with lower ionization parameter. However,
all of the other strong absorption features on \figcog seem to be consistent
with the curves of growth. One explanation for the discrepancy might be
an absorber component consisting of highly ionized dust. From
\S\ref{sec:overall} we have seen that the data are consistent with an
overabundance of neutral Fe, likely bound up in dust along the
line-of-sight. The dust that lies closest to the AGN may well be highly
ionized, contributing to a larger \oxyseven column than expected from a
simple single zone photoionization model. Certainly this could also
explain several highly ionized Fe features in \figufspeccont that are not
well fit by our best-fitting model in \S\ref{sec:modelfits}. Highly ionized dust can survive in warm, photoionized gas (see Reynolds \etal (1997) for further dicussion) although it seems probable in such scenarios that the warm material originates in the dusty cold material, which could be derived from a molecular torus.

\subsection{Detailed Comparison of Data and Photoionization Model}
\label{sec:comparison}

In this section we compare in detail the best-fitting XSTAR 
photoionized absorber model and the data (refer to 
\figufspeccont for the spectrum, and to \figvelprofone and \figvelproftwo
for some velocity spectra). Note that although the best-fitting
model was derived using spectra binned at $0.04 \AA$, we can compare
the model and data at higher spectral resolution, depending on the
signal-to-noise of the feature in question.
The strongest absorption-line features in the \chandra data are listed in Table~1.
\figufspeccont shows that the best-fitting XSTAR model described above
($N_{H} = 5.37 \times 10^{21} \ \rm cm^{-2}$, $\log{\xi}=2.52$ ergs cm $\rm{s}^{-1}$ )
appears to underpredict the strength of several absorption lines. However, 
as outlined in \S\ref{sec:modelfits}, the equivalent widths of lines in the 
XSTAR spectra do not correspond to a ($b$-value) appropriate for the data. For example, \sithirteen (r) \resonetwo ($\lambda 6.648$) in \figufspeccont shows a large discrepancy between the model and the data, but Table~1 does not.
A proper comparison between data and model for the stronger lines is 
described below. 

The model velocity profiles in \figvelprofone (for four of the
strongest absorption lines: \oxla, \nela, \mgla \& \sila were calculated 
using the appropriate $b$-value for the data (200 km $\rm{s}^{-1}$)
and show reasonable agreement between data and model.
In this case the model curves were calculated using the
broken power-law and two-edge continuum described in \S\ref{sec:overall},
a Gaussian for the absorption line, with 
$\sigma = b/\sqrt{2} = 141.4$~$\rm km \ s^{-1}$,
and the predicted model equivalent width from Table~1.
\figvelproftwo, which shows the velocity profiles of some of the 
strongest resonance absorption features and the profile of a transition 
in highly ionized Fe, illustrates that the best-fitting XSTAR 
model under-predicts some features. However the lower limits on 
the best-fitting column density and ionization parameter at 99$\%$ confidence, yield ionic columns that
give velocity profiles that are more consistent with 
the data for \neniner \resonetwo 
($\lambda 13.447$) and \mgelevenr 
\resonetwo ($\lambda 8.421$) (dashed lines in \figvelproftwo), without significant disimprovement of the profiles in \figvelprofone. Thus, in spite of the likely 
wide range of ionization states in the warm absorbing gas, a single-component 
photoionization 
model can give surprisingly good fits to the principal absorption lines.

There are many more discrete absorption features apparent in the
spectra but with lower signal-to-noise.
Although these individual weak features cannot be studied in 
detail we can at least check whether the best-fitting XSTAR model 
does not conflict with the data. We found that, as far as
narrow absorption lines are concerned, most absorption lines
predicted by the model were consistent with the data. In particular, 
we found that the model agrees 
well with the data for the low-order Lyman and He-like resonance transitions 
of Ne and O, although there may be blending associated with higher order 
transitions such as \nenine \resonefour ($\lambda 11.000$). 
Conversely, the data show possible strong absorption features which
are not predicted by the model. These features are mainly in the 
$\sim 1.1-1.5$ keV range (e.g. see \figufspeccont~(d)). Some of the strongest 
features may correspond to blends of L-shell 
transitions in very highly ionized Fe (\fetwenty-\fetwentyfive). The atomic 
database used by XSTAR 2.1.h does not yet include all transitions in highly 
ionized Fe and this would certainly explain these discrepancies between data 
and model.

The Fe ionic columns predicted by the best-fitting XSTAR 
model are largest for ionization states \fenineteen to \fetwentyfour. According to 
Behar, Sako, \& Khan (2001), an Fe~M-shell unresolved transition array
(UTA) due to Fe~{\sc i}-Fe~{\sc xv} lies between 17.5$\AA$ and 15$\AA$. According to 
Gu \etal (2001), there is a strong L-shell signature due to \fetwenty - 
\fetwentyfour between 10.5$\AA$ and 12.5$\AA$. Thus, some of the strong absorption
features between 0.85-1.0 keV in \figufspeccont may be part of, or blended with, 
an Fe UTA due to highly ionized Fe (our Fe ionic columns are largest for 
\fenineteen - \fetwentyfive). UTAs due to less highly ionized Fe (in the 
$\sim$ 15-17$\AA$ range) have been detected in some other Seyfert~1 galaxies 
(IRAS~13349$+$2438, Sako \etal 2001a; MCG~$-$6$-$30$-$15,  
Lee \etal 2001; NGC~3783, 
Kaspi \etal 2002; NGC~5548, Kaastra \etal 2002; Mkn~509,
 Yaqoob \etal 2003). 

In \S\ref{sec:overall} we mentioned how spectral fits 
using simple absorption-edge models (representing O~{\sc vii} 
and O~{\sc viii}) could be misleading.
We see that more sophisticated modeling with better data
also gives a weak O~{\sc vii} edge, as obtained from edge-fitting, 
but also a much weaker O~{\sc viii} edge since the 
data above the O~{\sc viii} edge are affected
by other absorption features (Fe UTAs and
 \neniner ($\lambda 13.447$) in particular) 
which can be accounted for by the photoionization models. 
This has also been pointed out by Lee \etal (2001).
The complex of \oxyseven $1s^{2}-1snp$ resonance 
transitions with $n>5$ may also complicate the region just above
the O~{\sc vii} edge (see also Lee \etal 2001). From the XSTAR models we can 
extract the \oxyseven and \oxyeight ionic column densities
at the extremes of the $90\%$ confidence intervals for $N_{H}$ versus $\xi$.
We find that the
\oxyseven column is $4.0^{+2.6}_{-1.8}\times 10^{15} \ {\rm cm^{-2}}$
and the \oxyeight column is
$2.7^{+1.2}_{-0.8}\times 10^{17} \ {\rm cm^{-2}}$.
Given an absorption cross-section at threshold of
$2.4 \times 10^{-19} \ \rm{cm^{-2}}$ for \oxyseven and $9.9 \times 10^{-20} \ \rm{cm^{-2}}$ for \oxyeight (Verner \etal, 1996), we find that the model optical depths
(at threshold) of the O edges are
$\tau_{{\rm \oxyseven}}=0.001^{+0.0006}_{-0.0005}$ and
$\tau_{{\rm \oxyeight}}=0.026^{+0.013}_{-0.007}$ respectively. 
These values can be compared with those obtained from an \asca
observation of \ngc in 1994, in which simple `edge-models' yielded
$\tau_{{\rm \oxyseven}}=0.26 \pm 0.04$ and
$\tau_{{\rm \oxyeight}}=0.09^{+0.04}_{-0.03}$
(\cite{reynolds97}; see also George \etal 1998).
Whilst the \asca result is inconsistent with our inferred 
\oxyseven optical depth, 
it \emph{is} consistent with the optical depth of our best fit neutral Fe 
L edge ($\tau=0.19 \pm 0.04$). Note that \asca 
resolution at energies around 0.7 keV is too coarse to resolve the two edges 
separately (see discussion in \S\ref{sec:overall}). From an Fe L3 edge depth of $0.19 \pm 0.04$, and given an Fe 2p absorption cross-section of $1.25 \times 10^{-18} \rm{cm}^{-2}$ (Verner \etal 1996), we infer a neutral Fe column of 
$1.5 \pm 0.3 \times 10^{17} \ \rm{cm}^{-2}$. An Fe~L1 (2s) edge (rest energy of 0.845 keV) optical depth of $\tau<0.05$ is allowed by the data, implying a neutral Fe column of $<3.6\times 10^{17} \rm{cm}^{-2}$ which is consistent with the estimate from the Fe~L3 edge, although we caution that the spectrum at both the Fe~L1 and the Fe~L2 (rest energy of 0.720 keV) edge energies is complex. Indeed, we cannot derive a  reasonable constraint on the Fe~L2 edge optical depth because of apparent emission blueward of this edge (see \figufspeccont). The implied neutral column density for cosmic abundances is then
$N_{H} \sim 3.8 \times 10^{21} \ {\rm cm}^{-2}$, which would act to extinguish 
the soft X-ray spectrum. Therefore the neutral Fe could
take the form of neutral dust, as argued by Lee 
\etal (2001). Our estimate of the \oxyeight optical depth from the 
photoionization modelling is inconsistent with both the \asca result and our 
direct fit of the \oxyeight edge ($\tau=0.14^{+0.04}_{-0.03}$). However, the XSTAR model fits the data well, so simple edge models may give the wrong optical depth. Nevertheless, we find that the measured equivalent widths of the \oxysevenr \resonetwo ($\lambda 21.602$) and \oxla absorption lines in Table~1 yield ionic column densities that are consistent with the edge optical depths measured in \S\ref{sec:overall}.

We investigated the robustness of the XSTAR model fits to the absorption
features by examining the data versus model
at the extremes of the $N_{H}$ versus $\xi$
confidence intervals (see \figxstarcontour). 
By picking pairs of values of $N_{H}$ and $\xi$ in the
plane of the contours, we can freeze the column density and
ionization parameter at the chosen values and find the
new best-fit for each pair of values (this is in fact
how such contours are constructed). Then we can examine
the data and model spectra in detail for each of these
new fits to see where the model and data deviate 
significantly from each other.
We found that 
the models were still able to account for the
strengths of all of the strong absorption-line features
(listed in Table~1) for pairs of
$N_{H}$ and $\xi$ lying inside the 99$\%$ confidence contour.
On the other hand, for pairs of  $N_{H}$ and $\xi$ 
lying outside the 99$\%$ contour we found
that absorption lines in the 12-15$\AA$ range were particularly
sensitive and the equivalent width would be over-predicted
or under-predicted by as much as a factor of two.
Therefore, we conclude that the 99$\%$ confidence contours in \figxstarcontour
are a good guide to the allowed range of $N_{H}$ and $\xi$.

As well as absorption,
the XSTAR code also calculates emission spectra. However, the 
very weak \neninef and \oxysevenf forbidden lines that we observe (see 
\figtriplets) cannot constrain the XSTAR emission spectra. Therefore our 
best-fit photoionization model of the warm absorbing gas (\figxstarmodel) 
does not warrant inclusion of the corresponding (weak) emission spectrum. 
Stronger emission
features have been detected in other observations of \ngc and we shall 
discuss potential constraints on the emission region from previous 
observations in \S\ref{sec:compare} below. The possibility of very broad 
relativistic emission lines in our soft X-ray data shall also be discussed 
in detail in \S\ref{sec:dust}.

\section{Comparison with previous observations}
\label{sec:compare}

Over the past decade \ngc has been observed in X-rays, by \asca on 1994 January 9, \bsax from 1997 December 31 to 1998 January 2 and most recently,
 by \xmm and \chandra \letg (non-simultaneously on 2000 July 2 and 2001 February 16 respectively). 
We measured a 2--10 keV flux of 
$4.0 \times 10^{-11}$ ergs $\rm{cm}^{-2}$ $\rm{s}^{-1}$ from \ngc with 
\chandra \hetg. This is comparable with the flux detected by \bsax (Guainazzi 
\etal 1999) and \asca (Reynolds 1997). Using Fig.1 of Kaastra \& Steenbrugge 
(2001), we find compared to the \chandra \letg and \xmm observations, our MEG flux at $20\AA$ is a factor $\sim 1.6$ and $\sim 1.9$ higher respectively. We compared our soft X-ray measurements 
with those from the \asca observation in \S\ref{sec:overall} and shall not 
discuss the \asca results further. \bsax observed a similar 
underlying continuum to that observed by us with \chandra \hetg. 
When the \bsax continuum is fit with a simple 
powerlaw model ($\Gamma= 1.70 \pm 0.04$), the main deviations include an 
absorption feature at 
$\sim$ 0.7 keV and an apparent soft excess between $\sim$ 0.3--0.7 keV. 
Guainazzi \etal (1999) measured two edges whose threshold energies and 
optical depths are consistent with the \chandra Fe L and \oxyseven 
measurements.

A strong soft excess was also observed by \chandra \letg at energies less 
than $\sim 0.65$ keV, when the power law describing the \letg data at 
$1-10\AA$ is extrapolated to lower energies. The \chandra \letg data also 
exhibit a flux deficit between $\sim 0.65 - 1.2$ keV relative to 
the extrapolated powerlaw. Kaastra \& Steenbrugge (2001) claim that
the \chandra \letg spectrum exhibits few strong discrete emission or 
absorption features, unlike our data. We note from Fig.~1 in 
Kaastra \& Steenbrugge (2001) (which shows counts spectra versus observed 
wavelength between 18-23$\AA$), that the \chandra \letg spectrum does exhibit
some absorption features. There is weak absorption due to \oxysevenr \resonetwo 
($\lambda 21.602$), a weak \oxysevenf emission feature and some absorption due 
to highly ionized Fe (possibly a blend of \fetwentythree ($\lambda 21.400$)
and \fetwentyfour ($\lambda 21.383$) inner shell transitions). 
Lower statistics during the \chandra \letg
observation may explain the weakness of many such features. In particular,
Kaastra \& Steenbrugge (2001) note the weakness of any possible absorption lines 
from highly ionized C, N and O. They find from photoionized models that ionic 
columns of the H- and He-like ions of C, N, O are $<10^{16} \ {\rm cm^{-2}}$. 
These ionic column densities are significantly lower than the 
$10^{16}-2\times 10^{17} \ {\rm cm^{-2}}$ range we find for H-like C, N, O ions 
from our best-fit model in \S\ref{sec:modelfits}. 
Kaastra \& Steenbrugge (2001) 
report that the best-fit model for the warm 
absorber predicts continuum absorption edges of less than a few percent 
for \oxyseven and \oxyeight. This agrees with our 
upper limit on the \oxyseven edge depth of $<0.03$, but is inconsistent with 
our estimate of the \oxyeight edge depth ($0.14^{+0.03}_{-0.04}$).
Kaastra \& Steenbrugge 
(2001) argue that one explanation of the flux reduction in the 12-15$\AA$ band
in the \chandra \letg data is a highly ionized warm absorber with 
$N_{H} \sim 2.4 \times 10^{21} \ {\rm cm^{-2}}$ and $\log \xi \sim 2.54$ ergs cm $\rm{s}^{-1}$. 
This value of the ionization parameter 
is in excellent agreement with that of our bestfit model, although their column 
density is less than half the value we observe. 
The lack of strong edges in the \letg and \xmm 
RGS data, and the much lower column 
density of their preliminary best-fitting warm absorber model, according to 
Kaastra \& Steenbrugge (2001), suggest that significant changes occurred 
in the warm absorber in the $\sim$ 230 days between the \letg and 
\hetg observations.

The details of the continuum observed and spectral features detected during 
the short observation of \ngc with \xmm RGS are 
not discussed by Kaastra \& Steenbrugge 
(2001). We note however, from Fig.1 in Kaastra \& Steenbrugge (2001), 
that the RGS1 data appear to exhibit a relatively strong \oxysevenf emission 
feature, unlike the \chandra \letg data, and there 
are hints of a P Cygni profile of \oxla. Also from Fig.1 in Kaastra \& 
Steenbrugge (2001), there is weak absorption in the RGS1 data due to \oxysevenr \resonetwo 
($\lambda 21.602$) and highly ionized Fe. The presence of the \oxysevenf emission
line in the \xmm RGS1 data and its absence in the \chandra \letg data from a year 
later, suggests that the \oxyseven emission region lies within 1 light year of the
source of the continuum radiation. The \oxysevenf emission feature likely
 originates in a photoionized plasma and the reduction in its intensity 
over the year between the
\xmm RGS observation and the \chandra \letg observation is consistent with the 
halving of the mean flux from this source in that time.
  
\section{Dust or Relativistically Broadened Lines?}
\label{sec:dust}

In the MEG spectrum of \ngc we observe a sudden drop in continuum flux of 
$\sim 20\%$ at $\sim 0.7$ keV (apparent in Figs. 1, 2 and 3) which we identify 
with an Fe L3 edge. A similar edge-like feature has been observed at 
the same (rest-frame) energy in \mcg and Mrk~766 with 
\xmm RGS (Branduardi-Raymont \etal 2001 (hereafter BR01), 
Sako \etal 2001b, Mason \etal 2002) and independently 
with \chandra \hetg (Lee \etal 2001). BR01, Sako \etal (2001b) and 
Mason \etal (2002) argue that the edge-like feature in particular, and more 
generally the soft excess in some Seyfert 1 galaxies, can be explained by 
relativistically-broadened emission lines due to C~{\sc vi} 
($\lambda 33.736$), \nila and \oxla. Lee \etal (2001) argue that 
the edge-like feature is due to a combination of neutral Fe 
bound-free L-shell absorption edge due to neutral dust and higher order (n $\geq 5$)  
\oxyseven resonance ${\it  1s^{2}-1s np}$ absorption. In this section we shall consider our observations in the context of each interpretation. 

\subsection{Testing the relativistic emission model for \ngc}

Due to the MEG bandpass, we can only test for
relativistically-broadened \oxla emission in our \ngc data.
Firstly, we added an emission-line
model from a relativistic
disk around a Schwarzschild black hole (e.g. Fabian \etal 1989)
to the best-fitting photoionized absorber model (with a broken power law 
model of the continuum but \emph{with the Fe L3 edge removed}).
The inner and outer radii of the disk were fixed
at 6 and 1000 gravitational radii respectively, and the
radial line emissivity per unit area was a power law
with index $-2.5$. The line energy was fixed at
0.6536 keV in the \ngc rest frame. The best-fitting
inclination angle and equivalent width were $39 \pm 1$ degrees and
$18^{+8}_{-5}$ eV respectively. The blue wing of the Schwarzschild disk line 
exhibits a sharp drop and gives a good overall fit to the sharp drop in flux
at $\sim$ 0.7 keV (see \figufspeccont), although the apparent flux deficit 
above $\sim$ 0.7 keV in \figufspeccont is somewhat underpredicted. 
The C-statistic of 
the best-fit Schwarzschild disk line was worse than that for the best-fit Fe L3
edge model by $\sim 30$ for the same number of parameters. We note that if
a blackbody is used for the soft continuum instead of a powerlaw, we obtain 
the same inclination angle but a larger equivalent width ($24^{+4}_{-7}$ eV)
and a C-statistic which is still larger by 29 than the Fe-L edge model.

When we replaced the \oxla emission from around a Schwarzschild black hole 
with the corresponding emission from around a maximally spinning Kerr hole, we found that the 
C-statistic \emph{worsened} by $\sim 6$ relative to the Schwarzschild disk 
line, for the same number of parameters. 
The best fitting inclination angle and equivalent width were 
$77^{+10}_{-27}$ degrees and $11 \pm 7$ eV respectively. We assumed the same 
radial emissivity law but extended the inner radius down to 1.24 gravitational
radii. The Kerr disk line 
\emph{cannot fit the data around 0.7 keV} because the blue wing of the Kerr 
disk line, being strongly broadened, is less sharp and `edge-like' than 
that of the Schwarzschild disk line.

A problem with the disk line interpretation is that if the drop in flux occurs at the same energy in several sources (and there are now three examples of such), the implied disk inclination angle would have to be identical for all the sources within $1^{o}$. An obvious test of the neutral dust model would 
be to measure the O~{\sc i} column predicted by the inferred Fe~{\sc i} 
column if it were locked up in dust. Unfortunately, due to contamination, 
the soft X-ray quantum efficiency of \chandra ACIS has degraded, and produces
significant excess absorption due to neutral O. Neutral dust along the line-of-sight might also reasonably be expected to contain silicates. We tested the data for a neutral Si~{\sc i} edge and found that, with the current calibration, $\tau <0.06$ is allowed, although this is not significantly detected. This yields a Si~{\sc i} column of $<4.4 \times 10^{17} \ \rm{cm}^{-2}$ which is consistent with the inferred Fe~{\sc i} column of $\sim 1.5 \times 10^{17} \ \rm{cm}^{-2}$. Alternative 
mechanisms for obtaining an overabundance of Fe~{\sc i} along the 
line-of-sight to the AGN could include shattered stellar cores from an 
environment rich in 
supernova remnants or a progenitor gamma-ray burst associated with 
the host galaxy (Piro \etal 1999).

\section{Summary of Observational Results}
\label{sec:results}
 
We observed \ngc 
for $\sim 79 \ {\rm ks}$ with the \chandra \hetg on 2001 June 29,
simultaneously with \rxte.
The complex Fe-K line and Compton-reflection
continuum will be discussed elsewhere. Here we summarize the main results from the \chandra
\hetg soft X-ray spectrum, in view of
relevant constraints from \rxte.
 
\begin{enumerate}
\item{The 0.5--5.0 keV data can be modeled with a broken power-law 
intrinsic continuum plus a single zone warm absorber model 
(see below). \rxte was used to constrain the hard power-law 
index ($1.794$), yielding a soft X-ray index of $ 2.27^{+0.08}_{-0.07}$ and a 
break energy of $1.07^{+0.08}_{-0.07}$ keV  when self-consistently modeled 
with the photoionized absorber. During the observing campaign 
the 2--10 keV flux was 
$4.0 \times 10^{-11}$ ergs ${\rm cm^{-2} \ s^{-1}}$, typical of
historical values (the corresponding 2--10 keV luminosity was
$6.1 \times 10^{42}$ ergs $\rm{s}^{-1}$, for $H_{0}=70 \ \rm km \ s^{-1} 
\ Mpc^{-1}$ and $q_{0}=0$).}

\item{Below $\sim 2$ keV, the {\it observed} spectrum
shows considerable deviations from a smooth continuum,
due mainly to bound-free absorption opacity, and complexes
of discrete absorption features.
Particularly noteworthy 
are the Fe L3 and \oxyeight edges at
$0.708\pm 0.003$ keV and $0.868^{+0.007}_{-0.010}$ keV
respectively. Within the errors,
both values are consistent with their respective rest energies, and 
from the best-fitting edge models the
optical depths at threshold are $0.18 \pm 0.04$ for Fe L3
and $0.14^{+0.03}_{-0.04}$ for O~{\sc viii}. The optical 
depth of the Fe L3 edge is consistent with the depth ascribed 
by \asca to an \oxyseven edge, whilst our \oxyeight edge 
depth is in marginal agreement with the \asca measurement 
(\cite{reynolds97}; \cite{george98}). We note that 
\asca spectral resolution was incapable of distinguishing an Fe 
L3 and \oxyseven edge and that estimates of the \oxyeight 
edge depth are also complicated by nearby absorption features due,
for example, to Ne and Fe. The neutral Fe edge may be due to dust in the 
warm absorber and therefore we tested the data for neutral Si. We found that, with the current calibration, a neutral Si~{\sc i} edge with $\tau <0.06$ is allowed, which yields a Si~{\sc i} column that is consistent with the inferred Fe~{\sc i} column. The soft X-ray quantum efficiency of \chandra ACIS has degraded due to contamination, and produces significant excess absorption due to neutral O, so we could not deduce whether there was excess absorption due to neutral O in \ngc. An alternative model of the neutral Fe edge region due to relativistically broadened emission lines from an accretion disk yielded a significantly worse fit to the data than an Fe L edge.}
 
\item{On the next level of detail,
we detect strong discrete absorption lines from H- and He-like ions of 
O, Ne, Mg, Si as well as discrete absorption features due to highly ionized 
Fe~{\sc xix-xxv}. Only the
\nila, \oxla and \mgla lines appear to be marginally resolved 
(see Table~1). A marginally resolved absorption feature at $\sim 12.82 \AA$ is likely to be a blend of \fetwenty lines.
 The remaining strong, discrete absorption features are unresolved
(FWHM$<275$~$\rm km \ s^{-1}$ for \oxysevenr \resonetwo ($\lambda 21.602$), going 
up to FWHM$<2145$~$\rm km \ s^{-1}$ for the complex \sila absorption feature).
The absorption lines are consistent with a net
outflow velocity of $\sim -$ 140 $\rm km \ s^{-1}$ relative to systemic.
However, the detailed velocity profiles of the absorption features 
are not all the same for all ionic species.
The complexity and differences in profiles could be due
to the contribution from different, unresolved
kinematic components and/or 
blending and contamination from absorption/emission due to other
atomic transitions.}

\item{We also detect \oxysevenr \resonetwo ($\lambda 21.602$) and \nila 
in our local frame, possibly indicating absorption in our Galaxy or 
local group of galaxies due to a hot medium. This absorption is likely to be 
due to the same medium as that responsible for the local ($z=0$) absorption observed by Nicastro \etal (2002).}
 
\item{We have modeled the \chandra \hetg spectra using the 
photoionization code XSTAR 2.1.h.
The best-fitting ionization parameter and equivalent Hydrogen column density of the absorbing gas are
$\log \xi=2.52^{+0.06}_{-0.04}$ ergs cm ${\rm s^{-1}}$ and
$N_{H}=5.37^{+1.45}_{-0.79}\times 10^{21} {\rm cm^{-2}}$ respectively.
This best-fitting model gives a good fit to
the overall {\it observed} spectrum. It also is able to
model most of the principal local absorption features well
but underpredicts some absorption lines such as \neniner \resonetwo 
($\lambda 13.447 $) by up to a factor of $\sim 2$ in 
equivalent width, but these are likely cases of blending with lines of highly ionized Fe. A curve-of-growth analysis indicates that a
velocity width of $b \sim 200 \ \rm km \ s^{-1}$ is consistent with the data.
Our results are insensitive to reasonable variations in the unobserved UV 
part of the spectral energy distribution.}
 
\item{We detect weak emission features due to \oxysevenf and \neninef.
However, our data cannot constrain the distance of
the emitter from the ionizing source. Since we cannot assume that
the emitter and absorber have the same ionization state and/or
column density, time-resolved spectroscopy with better signal-to-noise
is required to address the
question of the location of the absorber and emitter.
Deducing the mass outflow of the X-ray absorber must also await
better data because the density and global covering factor are
unknown. However, we obtain upper limits on the electron density of the 
emitting material of $<8 \times 10^{10} \ \rm{cm}^{-3}$ from the 
ratios of the \oxysevenf and \oxyseveni line strengths.
 We note also from observations of \ngc with \xmm and 
\chandra \letg (see \S\ref{sec:compare}), that the \oxysevenf line emission 
is variable on a timescale less than a year. Therefore the 
\oxysevenf emission region appears to lie within 
1 light-year of the ionizing radiation source.}  
 
\end{enumerate}

\section{Conclusions}
\label{sec:conclusions}
We observed the Seyfert~1 galaxy \ngc with the \chandra high energy 
transmission gratings and measure strong absorption lines from He-like 
O, Ne, Mg, Si, H-like N, O, Ne, Mg, Si and highly ionized Fe~{\sc xix-xxv}. 
The offset velocity of the strongest absorption profiles are slightly blueshifted but the errors are consistent with the systemic velocity of \ngc. Many profiles hint at the presence of either multiple kinematic 
components or blending. Absorption features due to \nila, \oxla and \mgla as well as a blend of Fe~{\sc xx} features appear to be marginally resolved. The soft X-ray spectrum of \ngc is adequately described by a simple, 
single-zone photoionized absorber
with an equivalent Hydrogen column density of $5.37^{+1.45}_{-0.79} 
\times 10^{21} \ \rm{cm}^{-2}$ and an ionization parameter of 
$\log \xi= 2.52^{+0.06}_{-0.04}$ ergs cm $\rm{s}^{-1}$ although there remain some features which are not identified. Although the photoionized gas
almost certainly is comprised of matter in more than one ionization state 
and may consist of several kinematic components, data with better 
signal-to-noise ratio and better spectral resolution are
required to justify a more complex model. In emission we detect only 
weak forbidden lines of \neninef and \oxysevenf.
We identify a spectral feature at 
$\sim 0.707$ keV with a neutral Fe L edge, which is most likely due to neutral dust along the line-of-sight to \ngc. However, this is not the only interpretation of this feature and a search for neutral O absorption which would reasonably be expected from dust absorption is complicated by contamination of the \chandra acis CCDs. Neutral Si absorption is allowed by the data and corresponds to a neutral Si column density that is consistent with the inferred neutral Fe column. We detect \nila and \oxysevenr \resonetwo ($\lambda 21.602 $) absorption 
at $z=0$, due to a local, hot medium. The hot, local ($z=0$) medium responsible is likely the same hot medium responsible for the absorption seen by Nicastro \etal (2002).

The authors gratefully acknowledge support from
NASA grants NCC-5447 (T.Y.), NAG5-10769 (T.Y.), NAG5-7385
(T.J.T) and CXO grant GO1-2102X (T.Y., B.M.).
This research
made use of the HEASARC online data archive services, supported
by NASA/GSFC and also  
of the NASA/IPAC Extragalactic Database (NED) which is operated by 
the Jet Propulsion Laboratory, California Institute of Technology, 
under contract with NASA.
The authors are grateful to the \chandra and \rxte
instrument and operations teams for making these observations
possible, and to Tim Kallman for much advice on XSTAR.
The authors also thank Julian Krolik
for useful discussions and Sandra Savaglio for contributing to one of the figures.

\newpage
\begin{deluxetable}{lrrrrr}
\tablecaption{Spectral Lines in the Chandra HETGS Spectrum of \ngc}
\tablecolumns{5}
\tablewidth{0pt}
\tablehead{
\colhead{Line $^{a}$} & \colhead{EW (Data)}$^{b}$ & \colhead{EW (Model) $^{c}$}
& \colhead{Velocity $^{d}$ } & \colhead{FWHM} & \colhead{$\Delta$ C$^{e}$} \nl
& \colhead{(eV)} & \colhead{(eV)} & \colhead{($\rm km \ s^{-1}$)} & \colhead{($\rm km \ s^{-1}$)} & \nl}
\startdata

\nila & $0.71^{+0.16}_{-0.38}$ & 0.45 & $-45^{+130}_{-120}$ & $350^{+585}_{-120}$ & $-11.1$\nl

\rm{N}~{\sc{vii}} \rm{Ly}$\alpha$ $(z=0)$ & $0.32^{+0.32}_{-0.32}$ & $\ldots$ & $-15^{+210,f}_{-270}$ & $<2535$ & $-3.2$\nl

\oxysevenr \resonetwo ($\lambda 21.602$) & $0.66^{+0.12}_{-0.27}$ & 0.24 & $-125^{+60}_{-35}$ & $<699$ & $-11.0$\nl

\oxysevenr \resonetwo $(z=0)$ & $0.48^{+0.25}_{-0.42}$ & $\ldots$ & $+85^{+175,f}_{-105}$ & $1$(fixed) & $-5.0$\nl

\oxysevenf ($\lambda 22.101 $) $^{g}$ & $1.33^{+0.83}_{-0.77}$ & $\ldots$ & $+125^{+85}_{-125}$ & $<480$ & $-17.2$\nl

\oxla & $1.65^{+0.34}_{-1.08}$ & 1.57 & $-310^{+136}_{-92}$ & $740^{+480}_{-145}$ &$-32.7$\nl

\neniner  \resonetwo ($\lambda 13.447$) & $0.91^{+1.05}_{-0.79}$ 
& 0.51 & $+65^{+75}_{-320}$ & $<960$ &$-7.6$\nl 

\neninef ($\lambda 13.698$) $^{g}$& $1.14^{+0.66}_{-0.57}$ & $\ldots$ & $+60^{+120}_{-210}$ & $<1260$ &$-15.2$\nl

\nela & $1.44^{+0.48}_{-0.47}$ & 1.95 & $-85^{+100}_{-100}$ & $<585$ & $-28.7$\nl

\mgeleven \resonetwo ($\lambda 9.169$) & $1.34^{+0.60}_{-0.55}$ 
& 0.96 & $-255^{+155}_{-155}$ & $<875$ & $-16.8$\nl 

\mgla  & $2.06^{+0.80}_{-0.69}$ & 1.99 & $-75^{+165}_{-170}$ 
& $775^{+645}_{-755}$ & $-34.3$\nl

\sithirteen \resonetwo ($\lambda 6.648$) & $2.66^{+0.63}_{-0.69}$ & 2.35 & $-255^{+110}_{-115}$ & $1$(fixed) &$-32.0$\nl

\sila  & $2.12^{+1.29}_{-1.09}$ & 2.52 & $-120^{+435}_{-555}$ 
& $<2145$ &$-10.2$\nl

\fetwenty ($\lambda 12.817 $) & $2.59^{+0.91}_{-0.77}$ & 1.72 & $+265^{+140}_{-140}$ 
& $996^{+455}_{-355}$ &$-50.8$\nl

\enddata

\tablecomments{\small Absorption-line parameters measured
from the MEG spectrum, using simple Gaussians
(see \S\ref{sec:features}).
All measured quantities refer to intrinsic parameters,
already corrected for the instrument response.
Errors are 90\% confidence for one interesting parameter
($\Delta C = 2.706$). All velocities have been rounded
to the nearest 5~$\rm km \ s^{-1}$.
$^{a}$ Laboratory-frame wavelengths. $^{b}$ Measured
equivalent widths in the \ngc frame. $^{c}$ Predicted equivalent widths
using $b=200 \ \rm km \ s^{-1}$ and 
XSTAR columns (\S\ref{sec:comparison}).
$^{d}$ Velocity offset (\ngc frame) of Gaussian
centroid relative to systemic. Negative values are blueshifts.
$^{e}$ Change in the fit statistic on addition of Gaussian.
$^{f}$ Velocity offset (galactic frame) of Gaussian
centroid relative to systemic.
$^{g}$ Emission line.
} 
\end{deluxetable}

\newpage

\begin{deluxetable}{lrlr}
\tablecaption{Ionic Column Densities ($10^{14} \ \rm{cm}^{-2}$)}
\tablecolumns{4}
\tablewidth{0pt}

\tablehead{
\colhead{Ion}  & \colhead{Predicted} & \colhead{Ion} & \colhead{Predicted} \nl  & \colhead{Column} &  &\colhead{Column}}

\startdata

H~{\sc i} &$2.3 \times 10^{0}$ & Si~{\sc xiii}  & $5.0 \times 10^{2}$\nl
C~{\sc iv} &$2.4 \times 10^{-5}$&Si~{\sc xiv}  & $9.3 \times 10^{2}$\nl
N~{\sc v} &$1.5 \times 10^{-4}$ &S~{\sc xv}  & $4.2 \times 10^{2}$\nl
O~{\sc vi} &$2.5 \times 10^{-2}$  &Fe~{\sc xviii}  & $9.6 \times 10^{1}$\nl
O~{\sc vii} &$4.0 \times 10^{1}$&Fe~{\sc xix}  & $3.2 \times 10^{2}$\nl
O~{\sc viii} &$2.7 \times 10^{3}$&Fe~{\sc xx}  & $5.4 \times 10^{2}$\nl
Ne~{\sc ix} &$8.2 \times 10^{1}$  &Fe~{\sc xxi}  & $2.8 \times 10^{2}$\nl
Ne~{\sc x} &$1.4 \times 10^{3}$ &Fe~{\sc xxii}  & $3.1 \times 10^{2}$\nl
Mg~{\sc xi} & $1.6 \times 10^{2}$&Fe~{\sc xxiii}  & $1.1 \times 10^{2}$\nl
Mg~{\sc xii}  & $8.5 \times 10^{2}$&Fe~{\sc xxiv}  & $2.2 \times 10^{1}$\nl

\enddata

\tablecomments{\small Column densities (in units of $10^{14} \ \rm{cm}^{-2}$) predicted by the best-fitting XSTAR model to the \chandra data (see \S\ref{sec:modelling}). The predicted column densities for Si~{\sc iii-iv} and C~{\sc ii-iii} from the XSTAR model were negligible.} 
\end{deluxetable}

\newpage

\section*{Figure Captions}
 
\par\noindent
{\bf Figure 1} \\
The \ngc MEG spectrum binned at 0.32$\AA$
compared with the best-fitting power-law model from a joint fit of \hetg and 3--8 keV \rxte data, extrapolated down to $0.5 \ {\rm keV}$.
 
\par\noindent
{\bf Figure 2} \\
Comparison of the \ngc MEG data (binned at $0.32\AA$) to 
a broken power-law model modified by two
absorption edges and Galactic
absorption ($1.97\times 10^{20} \ {\rm cm^{-2}}$). `Edge-models' have 
historically been used to model lower spectral resolution CCD data. 
The hard X-ray photon index is fixed at $1.794$, and
the best-fitting soft X-ray index and break energy
are $2.26\pm 0.05$ and $1.03^{+0.04}_{-0.03}$ keV respectively.
The best-fitting edge energies of 0.708$\pm$ 0.003 keV and $0.868^{+0.007}_{-0.010}$ keV agree well with the expected rest-frame energies
of the neutral Fe L3 (0.707 keV) and \oxyeight (0.871 keV) 
edges respectively. The addition of an \oxyseven edge (0.739 keV) does not significantly 
improve the fit. The threshold optical depth of the \oxyseven edge is $<$ 0.03 
at the $90\%$ confidence level. Note that using this model, 
the inferred optical depth of the \oxyeight edge could be
affected by the \nenine \resonetwo ($\lambda 13.447$) absorption feature 
(at $\sim 0.89 \ {\rm keV}$, observed).
The inferred depths of the \oxyseven and Fe L3 edges could also be influenced by 
complexity in the spectrum between $\sim 0.7-0.8$ keV,
in particular from Fe inner-shell and \oxyseven resonance absorption.

\par\noindent
{\bf Figure 3} \\
\ngc MEG observed photon spectrum (binned at $0.08\AA$) compared to the the best-fitting photoionized absorber model
(red solid line). Also shown is the intrinsic continuum
(blue solid line) modified by Galactic absorption
(neither the data nor model have been corrected for
Galactic absorption). The dashed lines show the expected positions
of some bound-free absorption edges.
The model consists of an
intrinsic continuum which is a broken power law (best-fitting
break energy at 1.07 keV), absorbed by photoionized gas
with best-fitting ionization parameter of $\log{\xi}$ = 2.52 ergs cm $\rm{s}^{-1}$ and column density $5.37 \times 10^{21}$ cm$^{-2}$.
Full details of the model calculations, fitting procedures,
and discussion of the details of the comparison between
data and model can be found in \S\ref{sec:modelling}.
The 0.7--1.1 keV region is very complex so simple
two-edge models fitted to older CCD data (of Seyfert~1 galaxies in general)
could have been biased by this complexity.
There is also an instrumental feature between $\sim$ 0.85--0.9 keV 
(see \figxstaroverdata) which may partly be responsible for the poor 
fit in this region. 
 
\par\noindent
{\bf Figure 4} \\
\chandra MEG photon spectrum for \ngc against
wavelength in the source rest-frame, using $z=0.0083$. The data are 
binned at $0.02\AA$, approximately the MEG FWHM spectral resolution (0.023$\AA$).
Labels show the Lyman series wavelengths (in blue) for Ar,
S, Si, Mg, Na, Ne,
O, and N. Also shown (in red) are the wavelengths of
the Helium-like triplets (resonance, intercombination and forbidden lines) of
Ar, S, Si, Mg, Ne and O.
 
\par\noindent
{\bf Figure 5} \\
Same spectrum as Fig. 4, but shown in blue are the wavelengths of He-like
resonance-absorption transitions ($n=1\rightarrow2,3,4,$..) of Ar, S, Si, Mg, 
Na, Ne, O and N. Also shown (in red) are K-shell transitions of \fenineteen in 
the wavelength range $2-25\AA$, as an illustration of some of the Fe blending 
in the data. 

\par\noindent
{\bf Figure 6} \\
\chandra MEG photon spectrum for \ngc against
observed wavelength in the vicinity of the He-like Ne and O triplets 
(binned at 0.04$\AA$ and 0.02$\AA$ respectively).  
Plotted in red are the wavelengths of the respective He-like
resonance, intercombination and forbidden transitions. 

\par\noindent
{\bf Figure 7} \\
Velocity profiles from combined \chandra MEG and HEG
spectra (except for \oxla which is from MEG data only since it is outside of the
bandpass of the HEG). The profiles correspond to
some of the strongest absorption features in the data (see Table~1).
The absorption features are well-described
by the column densities from the
best-fitting XSTAR photoionization
model (black solid lines), which has $\log{\xi} =2.52$ ergs cm $\rm{s}^{-1}$ and $N_{H}=5.37 \times 10^{21}$  $\rm{cm^{-2}}$
(see \S\ref{sec:modelling}). The model profiles were
calculated using the model EW and a Gaussian with $\sigma = b/\sqrt{2}$, for 
$b=200 \ \rm km \ s^{-1}$, inferred from a curve-of-growth analysis
(see \S\ref{sec:comparison} and \figcog). The profiles were given a uniform 
blueshift of $-140$ km $\rm{s}^{-1}$, corresponding to the weighted mean of the velocity shifts in Table~1, excluding the Fe~{\sc xx} absorption feature which is skewed due to likely blending (see \S\ref{sec:comparison}).
A velocity of zero corresponds to
the systemic velocity for \ngc (assuming $z=0.0083$).
 A negative velocity here indicates a blueshift relative to systemic.
The centroids of the absorption features lie between systemic and $\sim 
-$350 km ${\rm s^{-1}}$ (see Table~1), but all the profiles 
extend redward to the systemic velocity
and slightly beyond. FWHM velocity widths from
Gaussian fitting are given in Table~1 and can be
compared with 
the FWHM MEG velocity resolution
of 360, 565, 810 and 1105 km $\rm{s}^{-1}$ for \oxla, \nela, \mgla and 
\sila respectively, at observed wavelengths.

\par\noindent
{\bf Figure 8} \\
As Fig.7, for four more velocity spectra from combined \chandra MEG and HEG
data. Solid lines correspond to absorption profiles calculated from the 
best-fitting photoionization model as in \figvelprofone. Dashed lines 
correspond to absorption profiles calculated using the upper limits 
(at $99\%$ confidence) of the column density from the best-fitting 
photoionization model. Clearly the best-fit photoionization model 
under-predicts some of the features 
which the spectra are centered on (as labeled) but the upper limits on the 
ionic column densities are more consistent with the profiles for 
\neniner \resonetwo ($\lambda 13.447 $) and \mgelevenr \resonetwo ($\lambda 9.169$) in 
particular. The profiles were given a uniform blueshift of $-140$ km 
$\rm{s}^{-1}$, which corresponds to the weighted mean of the velocity 
offset of the strongest absorption features (see \S\ref{sec:comparison}).
FWHM velocities from
Gaussian fitting are given in Table~1 and can be compared with 
the FWHM MEG velocity resolution
of 535, 510, 745 and 1030 km $\rm{s}^{-1}$ for \fetwenty ($\lambda 12.817$), \nenine 
\resonetwo ($\lambda 13.447 $), 
\mgeleven \resonetwo ($\lambda 9.169 $) and \sithirteen \resonetwo 
($\lambda 6.648 $) respectively, at observed wavelengths.
Blending with Fe transitions is particularly likely to be associated 
with the \fetwenty ($\lambda 12.817$) and \nenine \resonetwo ($\lambda 13.447 $) transitions.
   
\par\noindent
{\bf Figure 9} \\
The observed and constructed
baseline SED (solid curve) used for photoionization
modeling of \ngc. 
Average radio and IR fluxes are from Ward \etal (1987).
The optical and UV data are fluxes from non-contemporaneous 
observations in NED. The X-ray portion is constructed from our \chandra \hetg 
and \rxte data. The hard X-ray power law has
$\Gamma=1.794$ and has been extended to 500 keV.
Given the significant variability in the UV/optical we have constructed upper and
lower envelopes enclosing the data.
The solid and dash-dot lines correspond to the upper (SEDa) and lower 
envelopes of 
optical-UV data in NED respectively. Note the prominent bump in the IR-UV 
region of the 
spectrum. The dashed line (SEDb) corresponds to simply joining the IR flux at 
0.33 eV and the X-ray flux at 0.5 keV with a straight line in log-log 
space and thus has no 
prominent bump in the IR-UV region. See \S\ref{sec:sed}
for full details of the construction and applications of the SEDs.
The dotted curve is the `mean AGN' SED of  Matthews \& Ferland (1987),
normalized to the same ionizing luminosity (i.e. in 
the range 1--1000 Rydberg) as that of the \ngc
baseline SED (SEDa). 

\par\noindent
{\bf Figure 10} \\
Comparison of MEG data (top panel, raw counts spectrum) binned at $0.16\AA$
with the best-fitting
XSTAR photoionization model (red curve; see \S\ref{sec:modelling})
and the ratio of data to model (bottom panel), showing a good overall
fit to the data. In this representation the various instrumental
edges and features can be seen, and in particular the poor fit around 2.0--2.5 keV
can be seen to coincide with the large jump in
the X-ray telescope effective area.
 
\par\noindent
{\bf Figure 11} \\
Best-fitting XSTAR (v.2.1.h) photoionization
model spectrum fitted to the MEG and HEG data as described in \S\ref{sec:modelling}
(see also \figufspeccont).
The ionization parameter is $\log{\xi}=2.52$ ergs cm $\rm{s}^{-1}$, and $N_{H} = 5.37 \times
10^{21} \rm \ cm^{-2}$. The intrinsic continuum is a
broken power law. A fitted neutral Fe edge at 0.708 keV, and 
Galactic absorption are also included. The purpose of this plot is to show the effects of the instrumental effective area, especially where it changes abruptly. Note that this spectrum is for $b=$1 
km $\rm{s}^{-1}$ so the 
equivalent widths of absorption features are limited by the energy resolution of
XSTAR and are not as accurate as those computed for \figvelprofone and \figvelproftwo (to which the reader should refer for details of the absorption features). Full details and other parameters
can be found in the text (in \S\ref{sec:overall} and \S\ref{sec:modelling}).
 
\par\noindent
{\bf Figure 12} \\
The $68\%, 90\%$, and $99\%$ confidence contours of
the logarithm of the ionization parameter ($\log{\xi}$) versus
the column density ($N_{H}$) around the best-fitting photoionized absorber
corresponding to \figxstaroverdata (see also \figufspeccont and 
\figxstarmodel).
The HEG and MEG spectra were binned at $0.04 \AA$.
See \S\ref{sec:modelling} for details.

\par\noindent
{\bf Figure 13} \\
Curves of growth for various values of the velocity width,
$b$. Plotted for each $b$ is the logarithm of line equivalent width (EW)
per unit wavelength against the logarithm of the product
of ionic column density ($N$), oscillator strength ($f$), and
wavelength ($\lambda$). 
The plotted points are taken from the measured equivalent widths (converted
to $\AA$ using the values in Table 1) and predicted ionic column densities
(from the best-fitting XSTAR model described in \S\ref{sec:modelling}). 
These values are consistent with $b=$200 km $\rm{s}^{-1}$ within errors.

\newpage

\begin{figure}[h]
\vspace{10pt}
\centerline{\includegraphics[width=8.0in,height=8.0in]{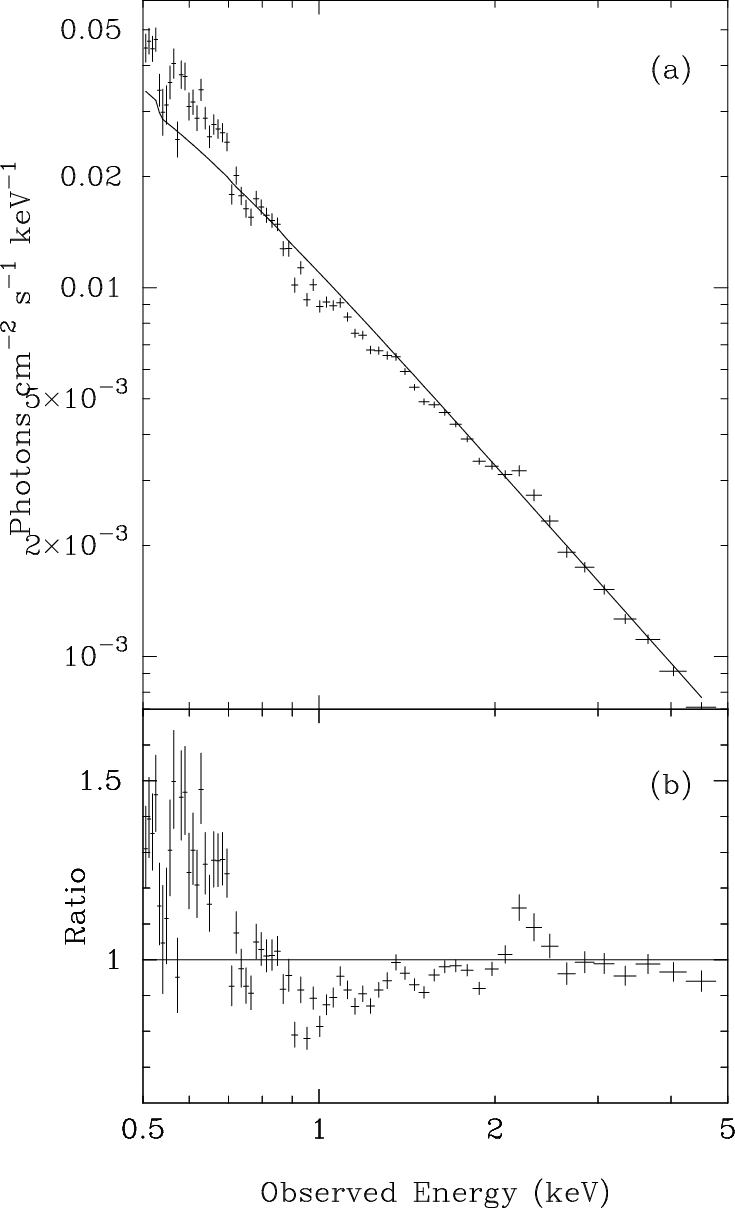}}
\caption{ }
\end{figure}

\begin{figure}[h]
\vspace{10pt}
\centerline{\includegraphics[width=8.0in,height=8.0in]{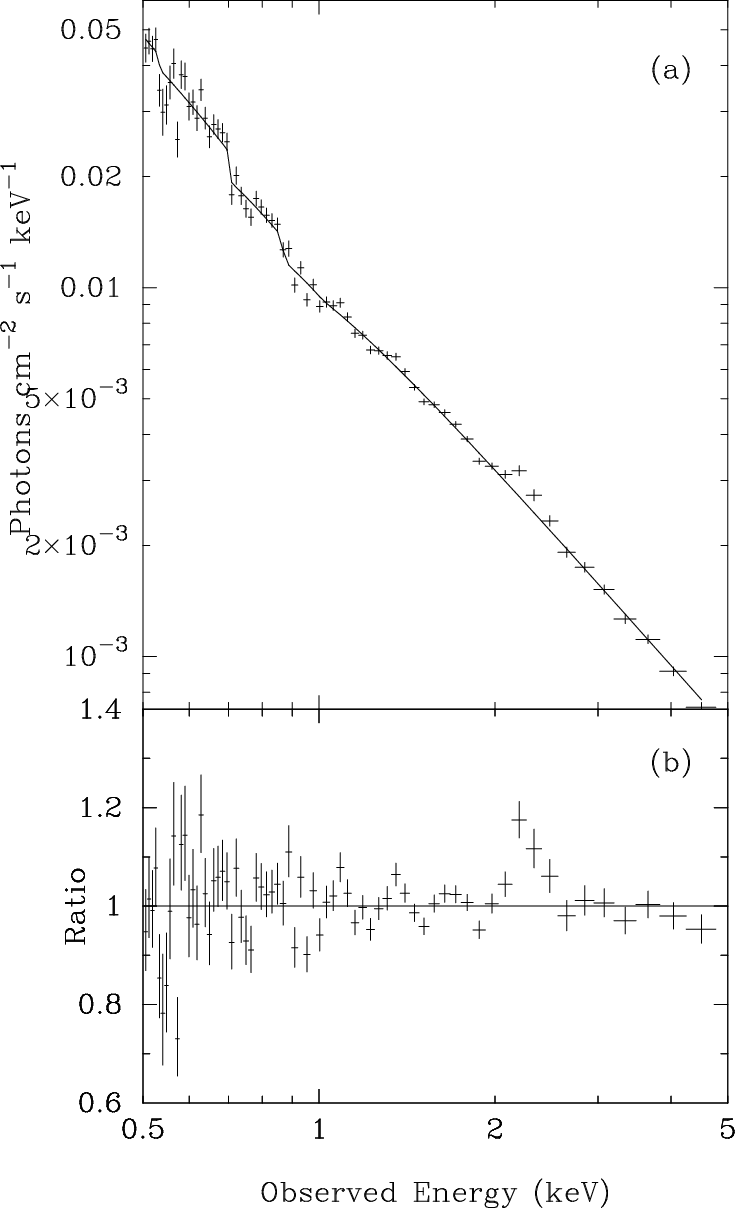}}
\caption{ }
\end{figure}

\begin{figure}[h]
\vspace{10pt}
\centerline{\includegraphics[width=8.0in,height=8.0in]{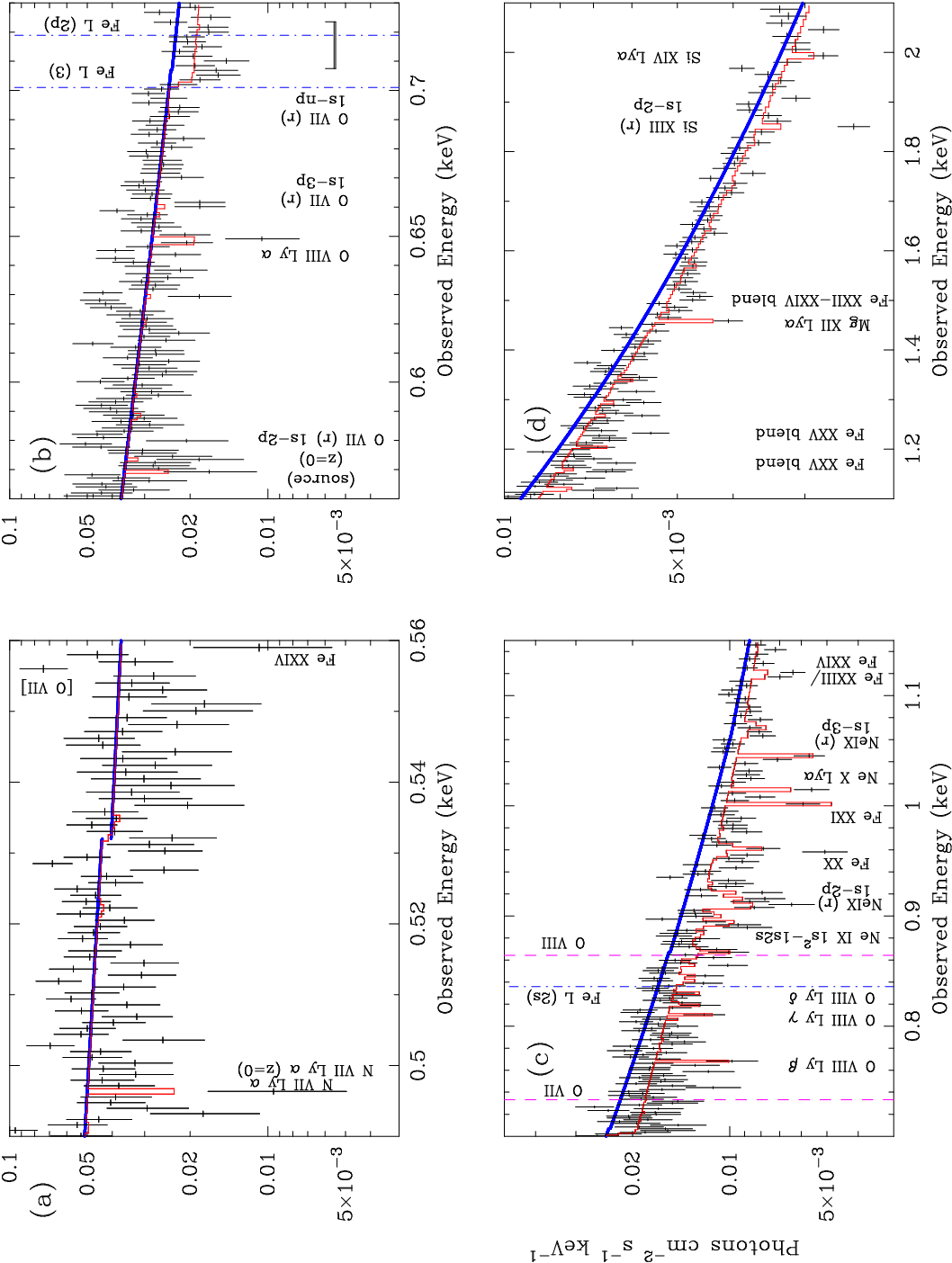}}
\caption{ }
\end{figure}

\begin{figure}[h]
\vspace{10pt}
\centerline{\includegraphics[width=8.0in,height=8.0in]{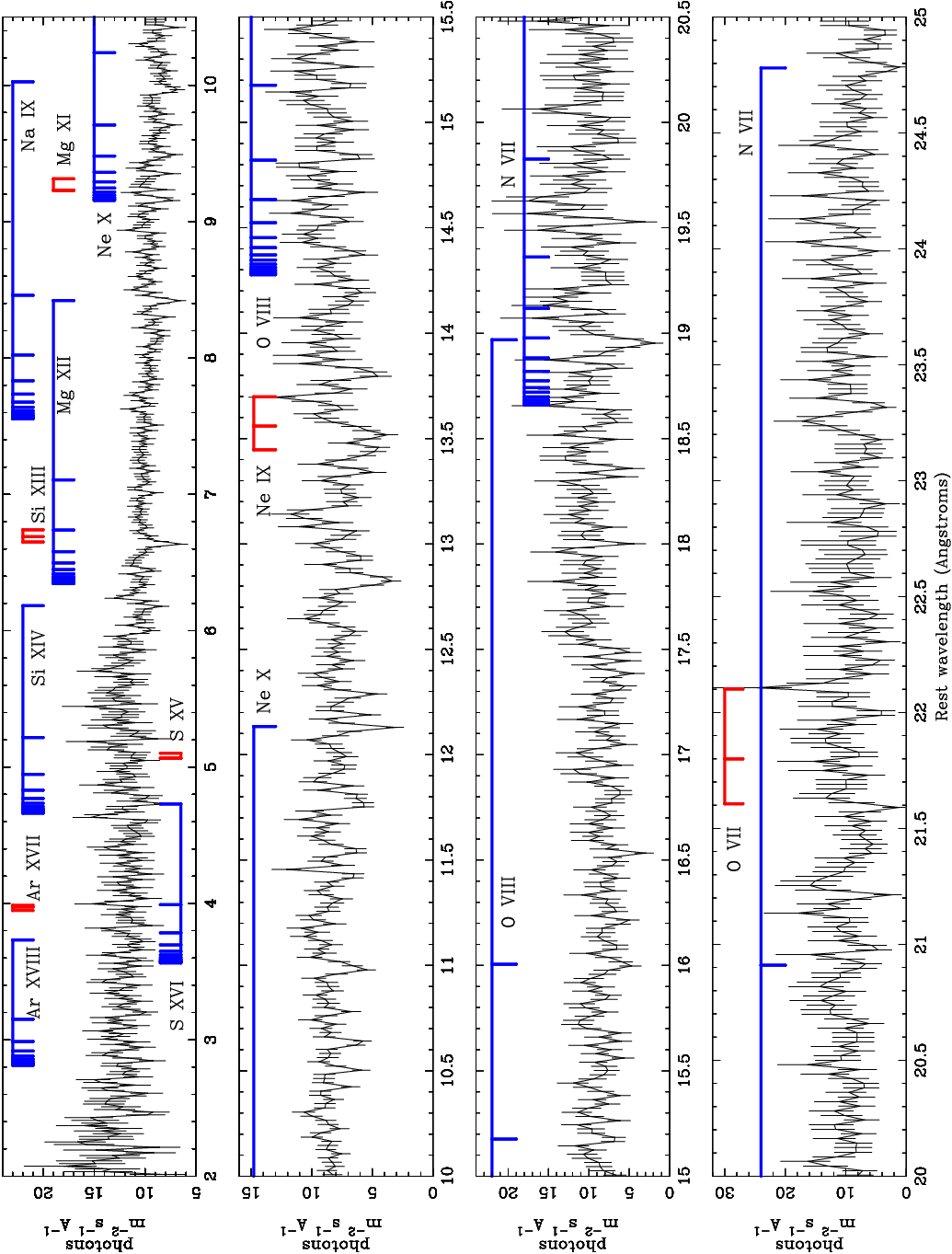}}
\caption{ }
\end{figure}

\begin{figure}[h]
\vspace{10pt}
\centerline{\includegraphics[width=8.0in,height=8.0in]{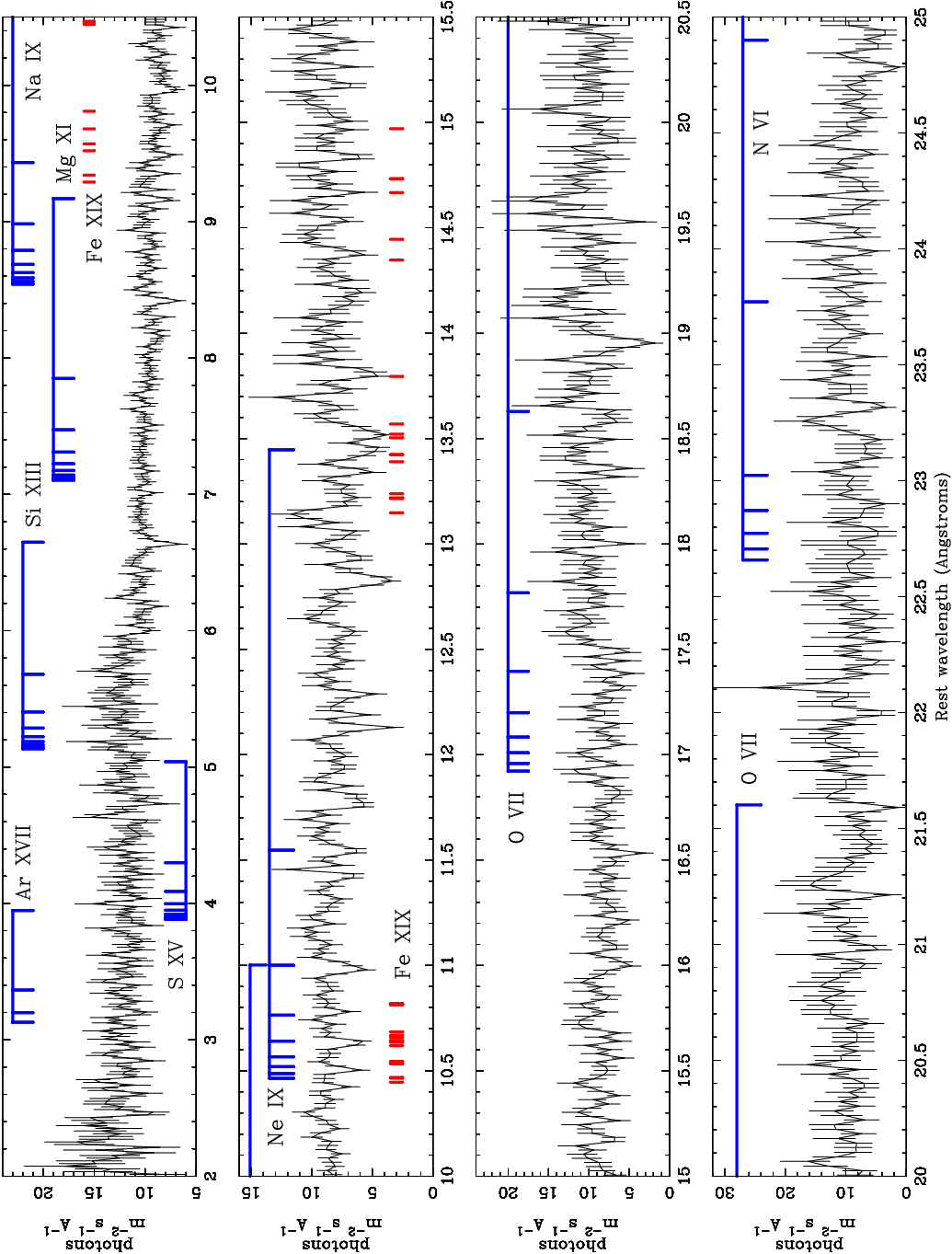}}
\caption{ }
\end{figure}

\begin{figure}[h]
\vspace{10pt}
\centerline{\includegraphics[width=8.0in,height=8.0in]{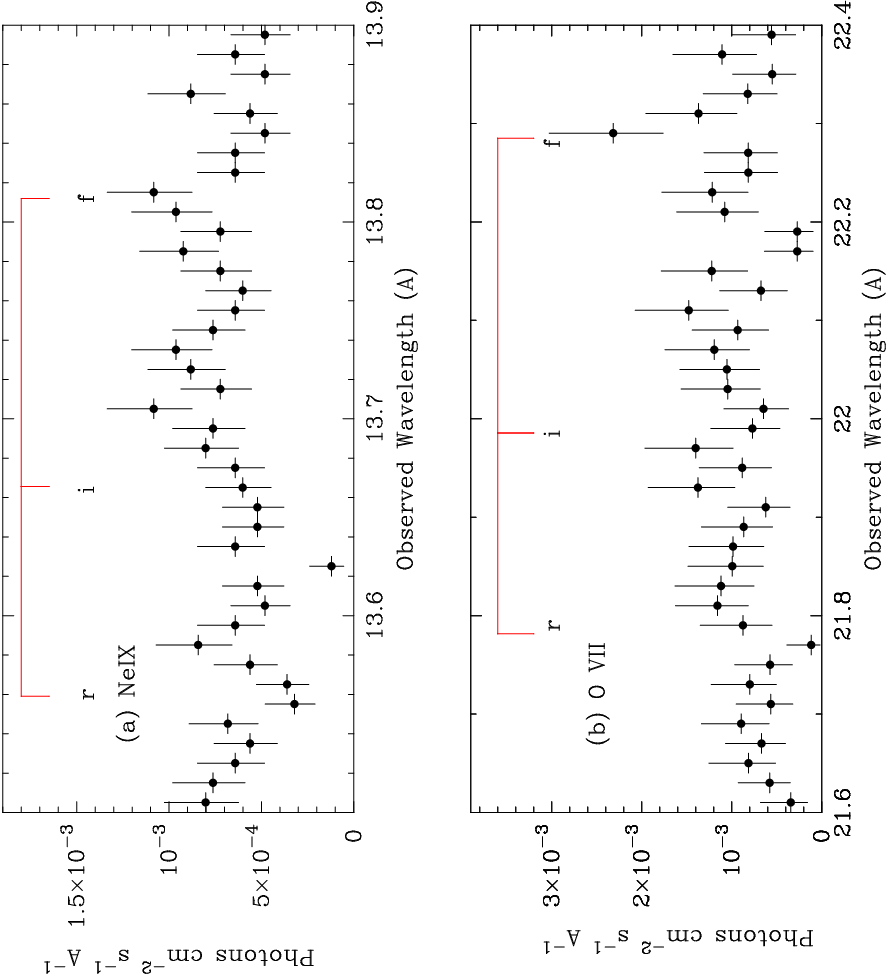}}
\caption{ }
\end{figure}

\begin{figure}[h]
\vspace{10pt}
\centerline{\includegraphics[width=8.0in,height=8.0in]{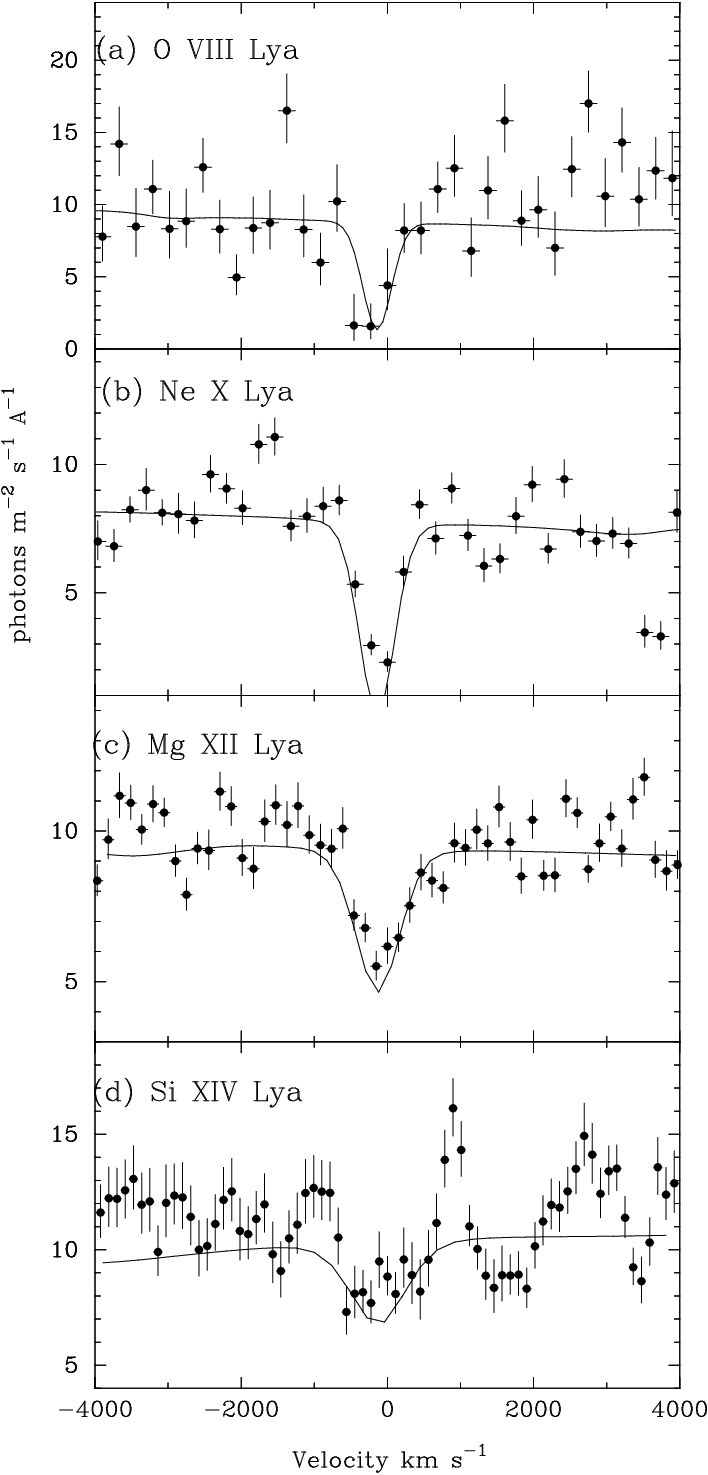}}
\caption{ }
\end{figure}

\begin{figure}[h]
\vspace{10pt}
\centerline{\includegraphics[width=8.0in,height=8.0in]{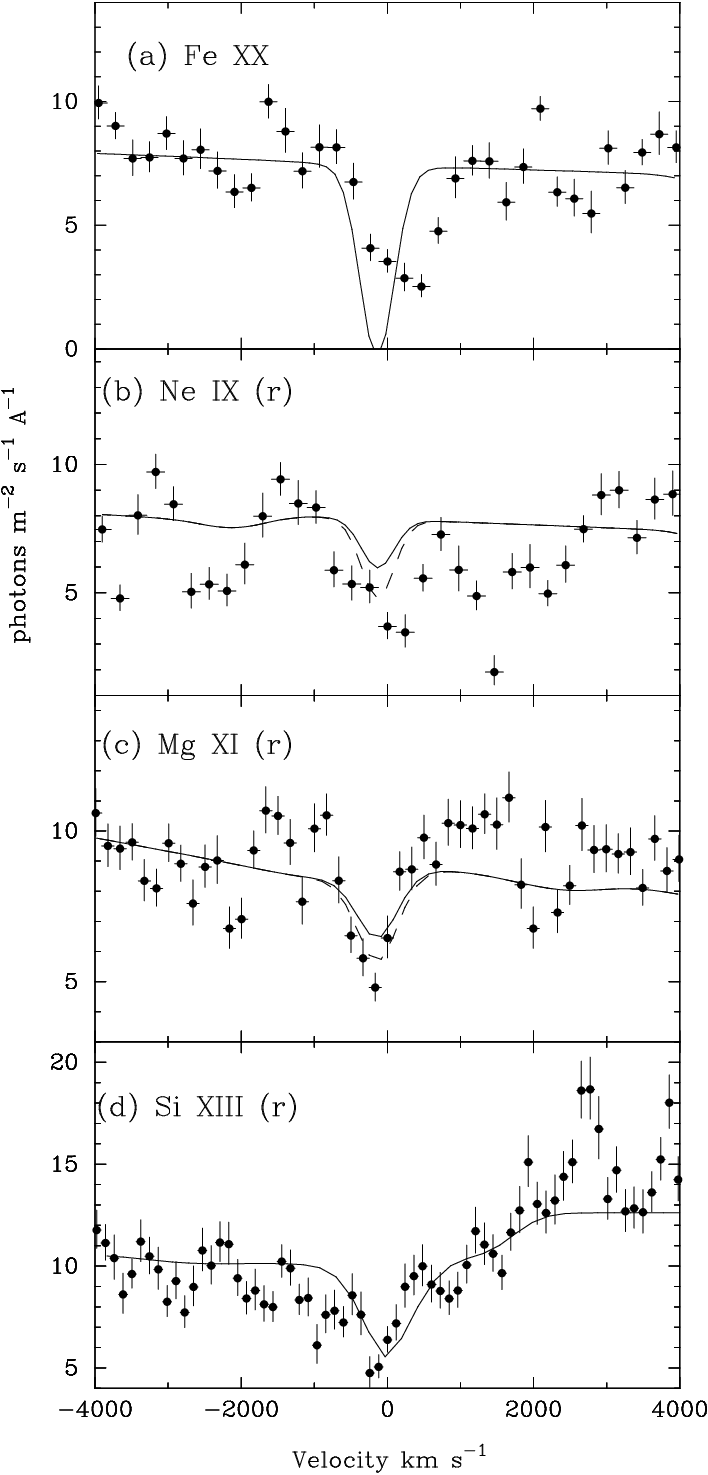}}
\caption{ }
\end{figure}

\begin{figure}[h]
\vspace{10pt}
\centerline{\includegraphics[width=8.0in,height=8.0in]{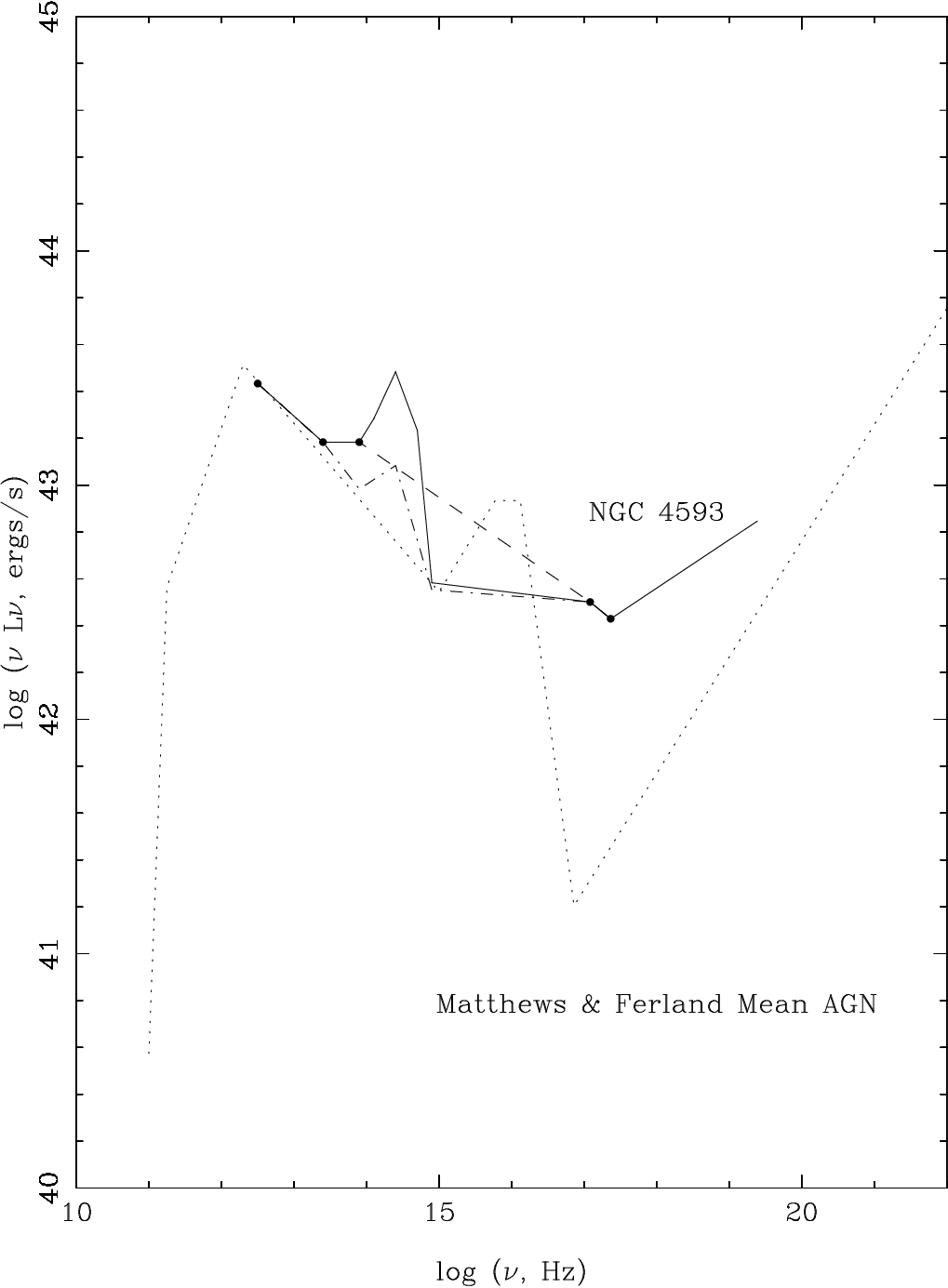}}
\caption{ }
\end{figure}

\begin{figure}[h]
\vspace{10pt}
\centerline{\includegraphics[width=8.0in,height=8.0in]{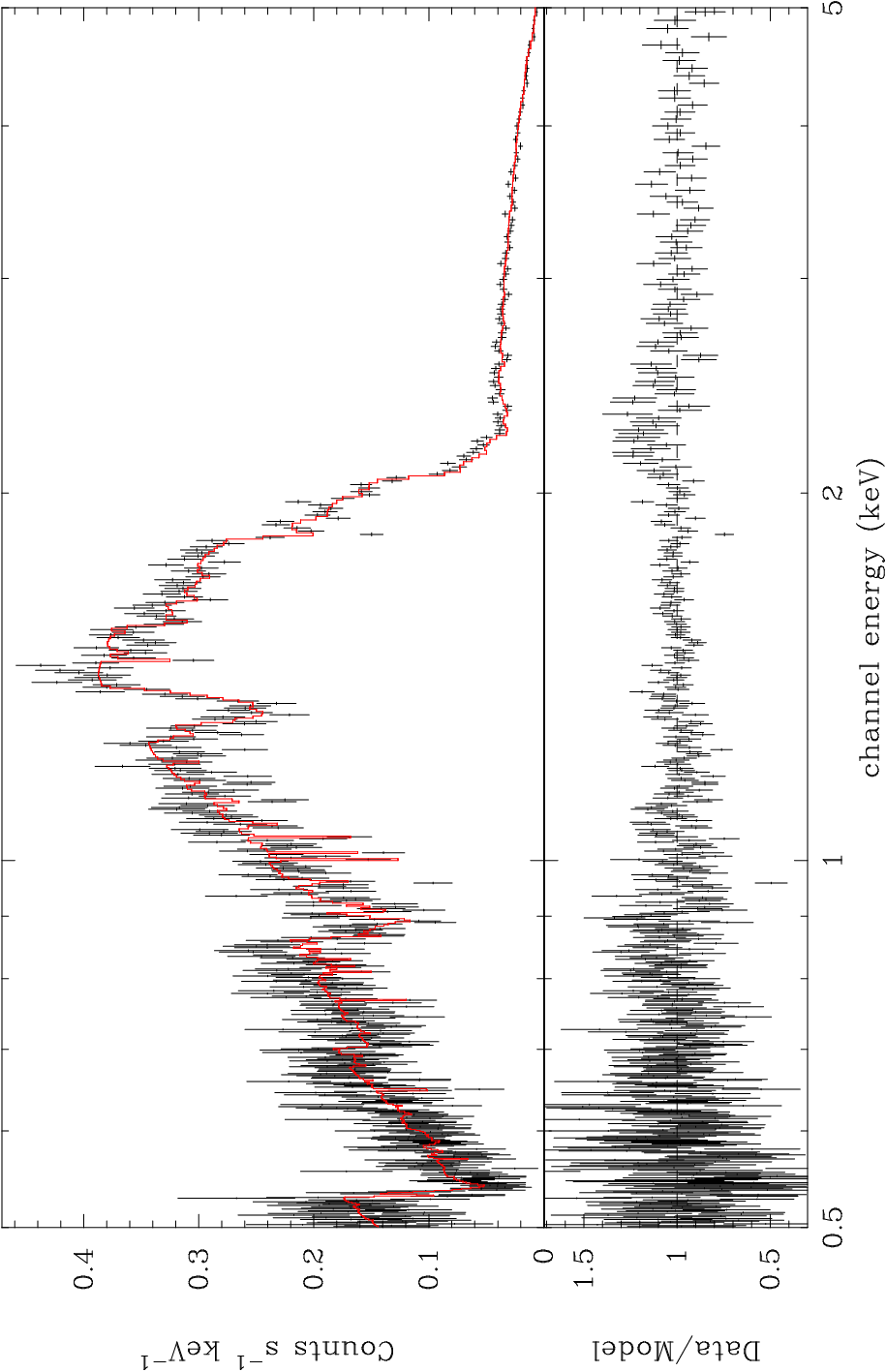}}
\caption{ }
\end{figure}

\begin{figure}[h]
\vspace{10pt}
\centerline{\includegraphics[width=8.0in,height=8.0in]{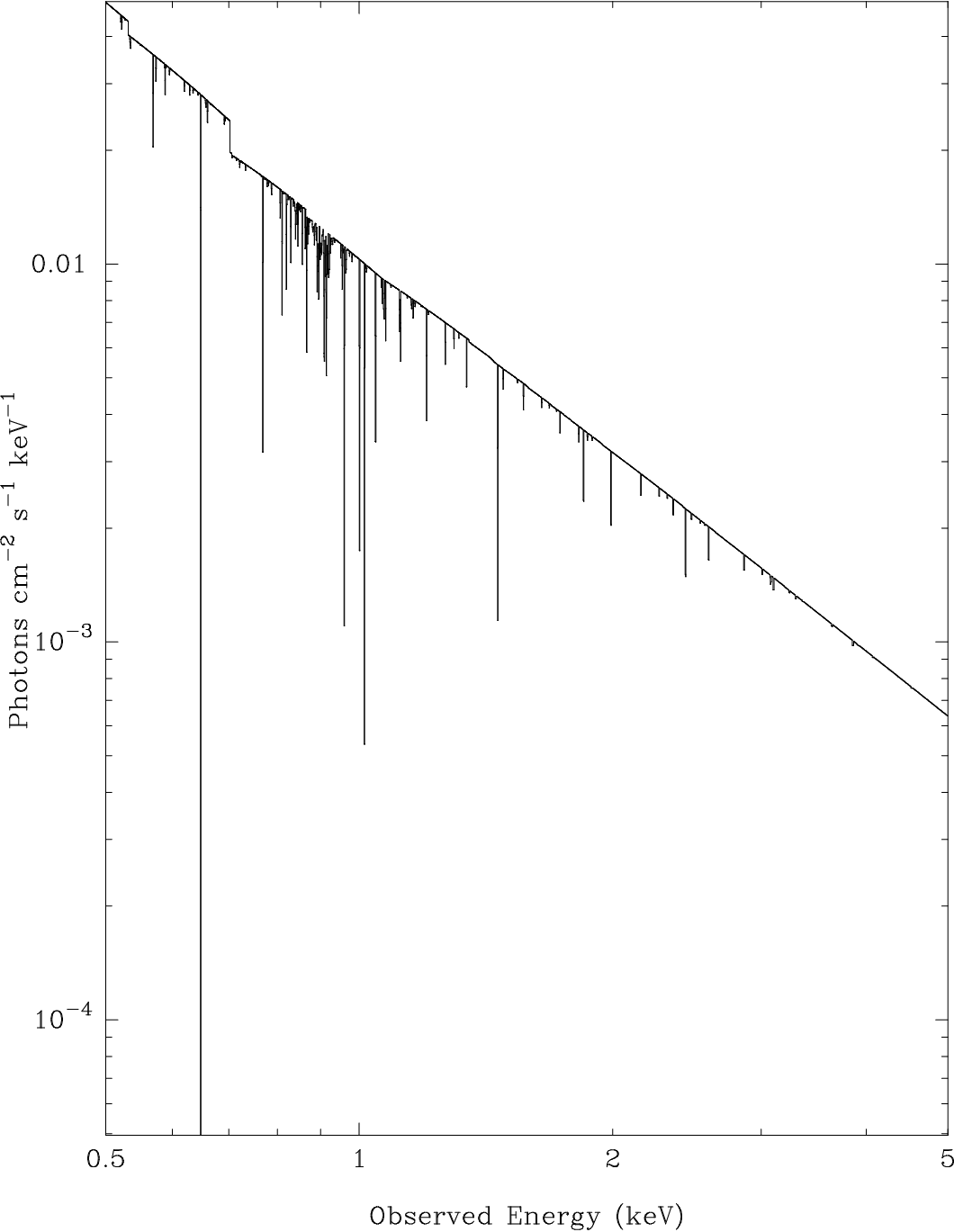}}
\caption{ }
\end{figure}

\begin{figure}[h]
\vspace{10pt}
\centerline{\includegraphics[width=8.0in,height=8.0in]{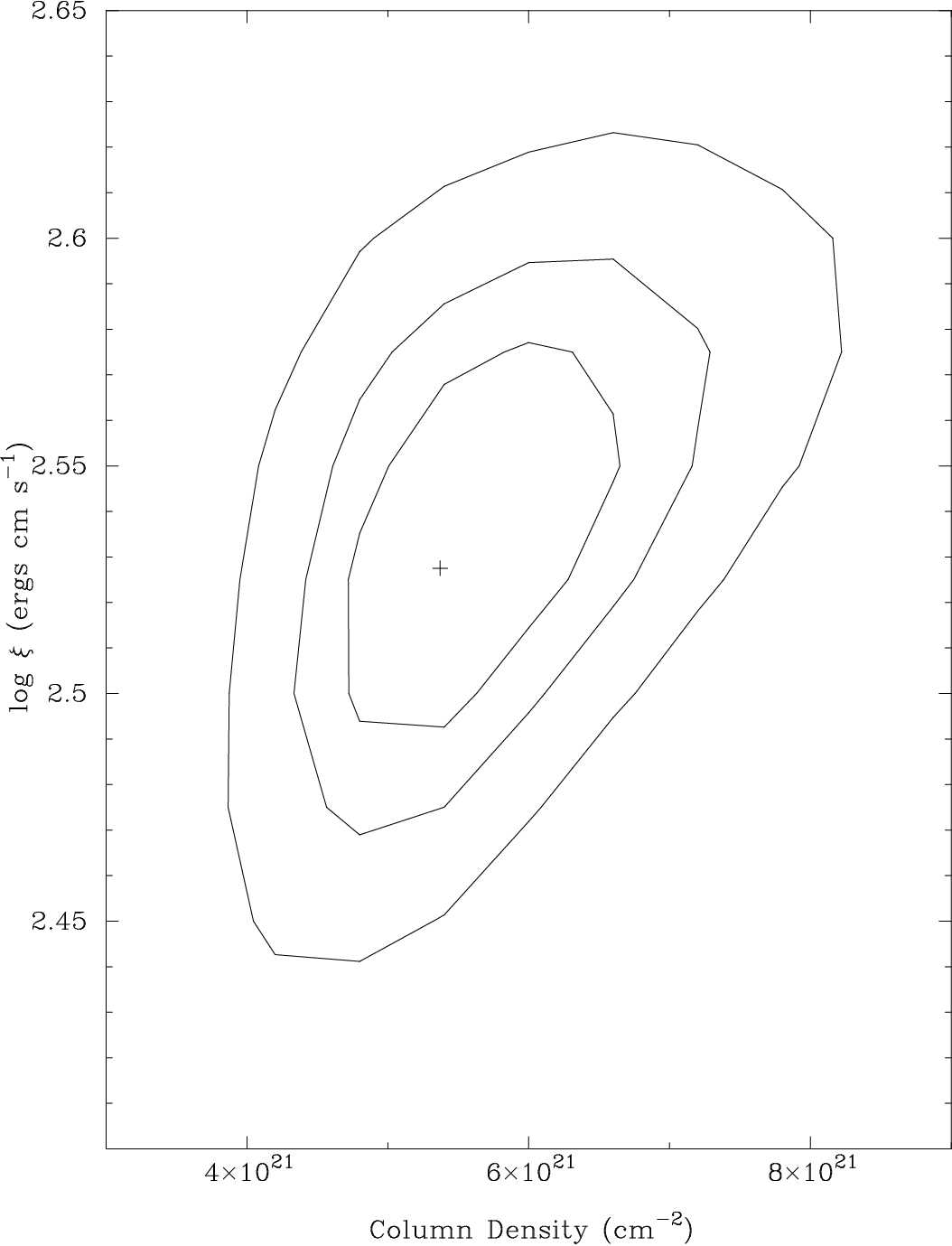}}
\caption{ }
\end{figure}

\begin{figure}[h]
\vspace{10pt}
\centerline{\includegraphics[width=8.0in,height=8.0in]{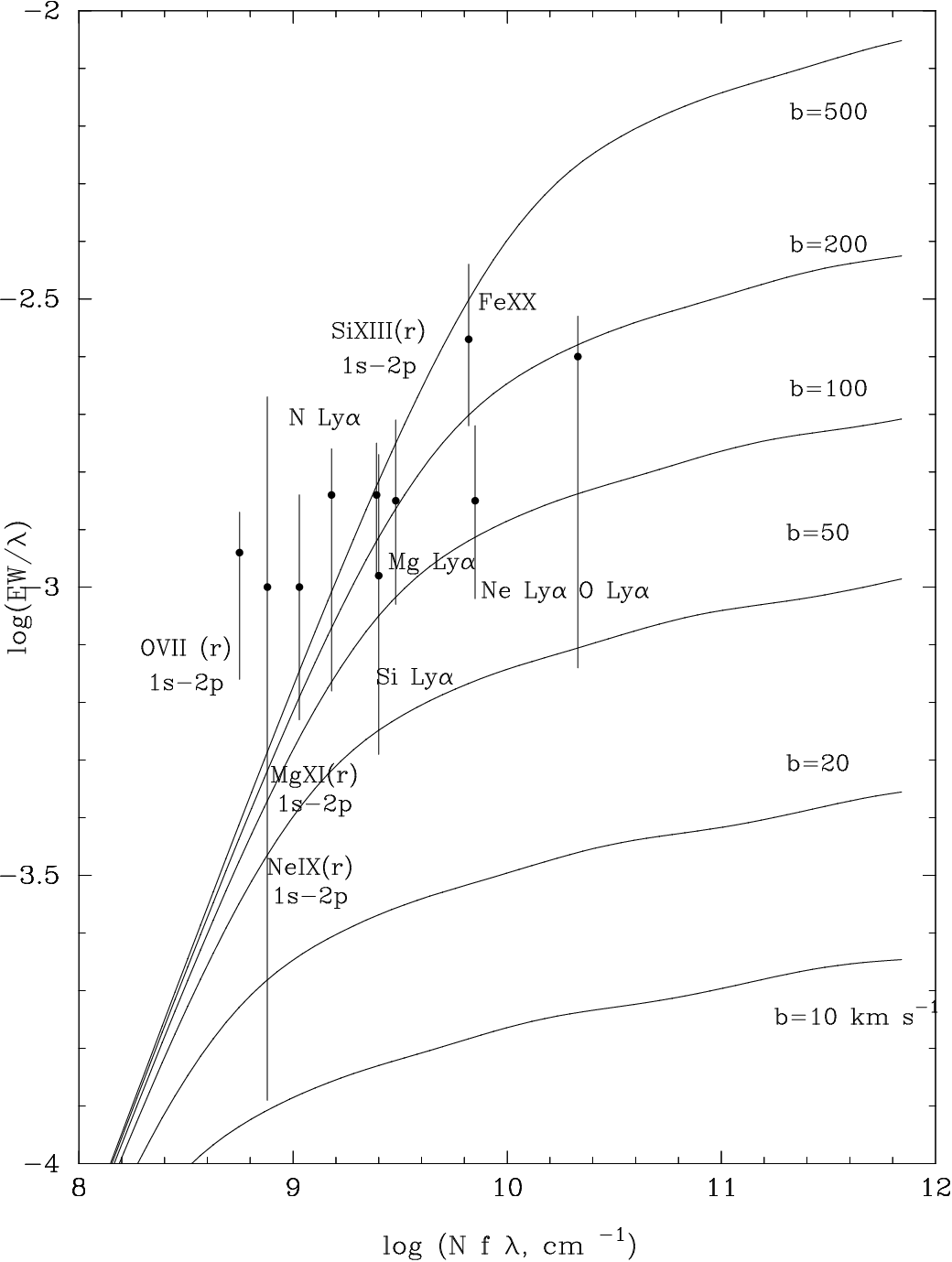}}
\caption{ }
\end{figure}


\newpage

\begin{thebibliography}{}

\bibitem[Behar \etal 2001]{behar01} Behar, E.,
Sako, M., \& Kahn, S. M. 2001, \apj, 563, 497

\bibitem[Branduardi-Raymont \etal 2001]{branduardi01} Branduardi-Raymont, G.,
Sako, M., Kahn, S. M., Brinkman, A. C., Kaastra, J. S., \& Page, M. J. 2001, 
\aap, 365, 140

\bibitem[Cash 1976]{cash76} Cash, W., 1976, \aap, 52, 307

\bibitem[De Vaucouleurs \etal 1991]{devauc91} De Vaucouleurs, G. \etal, 1991, 
3rd Ref. Cat. of Bright Galaxies, v 3.9

\bibitem[Elvis \etal 1989]{elvis89} Elvis, M., Wilkes, B. J., Lockman, F. J. 1989, \aj, 97, 777

\bibitem[Fabian \etal 1989]{fabian89} Fabian, A. C., Rees, M. J., Stella, L.,
\& White, N. E. 1989, MNRAS, 238, 729

\bibitem[Gehrels 1986]{gehrels86} Gehrels, N. 1986, \apj, 303, 336 

\bibitem[George \etal 1998]{george98} George, I. M., Turner, T. J., Netzer, H., Nandra, K., Mushotzky, R. F., \& Yaqoob, T. 1998, \apjs, 114, 73

\bibitem[Gu \etal 2001]{gu01} Gu, M. F., Kahn, S. M., Savin, D. W., Behar, E.,
Beiersdorfer, P., Brown, G. V., Liedahl, D. A., Reed, K. J., 2001, \apj, 563, 462  

\bibitem[Guainazzi 1996]{guain96} Guainazzi, M., Mihara, T.,
Otani, C., \& Matsuoka, M. 1996, PASJ, 48, 781

\bibitem[Kaastra \& Steenbrugge 2001]{kaasteen02} Kaastra, J. S., 
Steenbrugge, K. C., 2001, in Conf. Proc., X-ray Emission from Accretion onto Black Holes, Proceedings, ed. T. Yaqoob, \& J. H. Krolik (published electronically on ADS), E79 (astro-ph/0111419)

\bibitem[Kaastra \etal 2002]{kaastra02} Kaastra, J. S., 
Steenbrugge, K. C., Raassen, A. J. J., 
van der Meer R. L. J., Brinkman, A. C., Liedahl, D. A., Behar, E., 
\& de Rosa, A.  2002, A\&A, 386, 427 

\bibitem[Kaspi \etal 2002]{kaspi02} Kaspi, S., \etal 2002, ApJ, 574, 643 

\bibitem[Lee \etal 2001]{lee01} Lee, J. C., Ogle, P. M., Canizares, C. R., 
Marshall, H. L., Schulz, N. S., Morales, R., Fabian, A. C., \& Iwasawa, K. 2001, ApJ, 554, L13

\bibitem[Lee \etal 2002]{lee02} Lee, J. C., Reynolds, C. S., Remillard, R., Schulz, N. S., Blackman, E. G., \& Fabian, A. C. 2002, ApJ, 567, 1102

\bibitem[Markert \etal 1995]{mark1995} Markert, T. H., Canizares, C. R., Dewey, D.,
McGuirk, M., Pak, C., \& Shattenburg, M. L. 1995, Proc. SPIE, 2280, 168

\bibitem[Matthews \& Ferland 1987]{matthews87} Matthews, W. G., \& Ferland, G. J. 1987, 
\apj, 323, 456 

\bibitem[Mason \etal 2002]{mason02} Mason, K. O. \etal 2002, \apj, accepted (astro-ph/0209145)

\bibitem[Nicastro \etal 2002]{nicastro02} Nicastro, F. \etal 2002, \apj, 573, 157

\bibitem[Piro \etal 1997]{piro97} Piro, L., Matt, G., \& Ricci, R. 1997, \aap, 126, 525 

\bibitem[Piro \etal 1999]{piro99} Piro, L., \etal 1999, \apj, 514, L73 

\bibitem[Reynolds 1997]{reynolds97} Reynolds, C. S. 1997, \mnras, 286, 513

\bibitem[Reynolds \etal 1997]{reynolds97a} Reynolds, C. S., Ward, M. J., Fabian, A. C., Celotti, A.  1997, \mnras, 291, 403

\bibitem[Sako \etal 2001a]{sako01a} Sako, M., \etal 2001a, A\&A, submitted, 365, L168

\bibitem[Sako \etal 2001b]{sako01b} Sako, M., \etal 2001b, \apj, submitted (astro-ph/0112436)

\bibitem[Turner \etal 2001]{turner2001} Turner, T. J., \etal 2001, \apj, 548, L13

\bibitem[Verner \etal 1996]{verner96} Verner, D. A., Ferland, G. J., Korista, K. T. \& Yakovlev, D. G. 1996, \apj, 465, 487

\bibitem[Ward \etal 1987]{ward1987} Ward, M., Elvis, M., Fabbiano, N., 
Carleton, P., Willner, S. P., \& Lawrence, A. 1987, \apj, 315, 74

\bibitem[Weaver \etal 1998]{weaver98} Weaver, K. A., Krolik, J. H., \&
Pier, E. A. 1998, \apj, 498, 213 
 
\bibitem[Yaqoob \etal 2003]{yaqoob02} Yaqoob, T., McKernan, B.,  
Kraemer, S. B.,
Crenshaw, D. M., George, I. M., \& Turner, T. J. 2003a, \apj, 582, 105

\end{thebibliography}
\end{document}